\documentclass{article}\usepackage[]{graphicx}\usepackage[]{xcolor}
\makeatletter
\def\maxwidth{ %
  \ifdim\Gin@nat@width>\linewidth
    \linewidth
  \else
    \Gin@nat@width
  \fi
}
\makeatother

\usepackage{Sweave}

\usepackage[a4paper]{geometry}
\usepackage[authoryear]{natbib}
\usepackage{thumbpdf,lmodern}

\usepackage{amsmath}
\usepackage{amsthm}
\usepackage{amsfonts}
\usepackage{amssymb}
\usepackage{enumitem}

\usepackage[textsize=small]{todonotes}

\usepackage{natbib}

\usepackage{layouts}

\newcommand{\proglang}[1]{\textsf{#1}}
\makeatletter
\newcommand\code{\bgroup\@makeother\_\@makeother\~\@makeother\$\@codex}
\def\@codex#1{{\normalfont\ttfamily\hyphenchar\font=-1 #1}\egroup}
\makeatother
\newcommand{\pkg}[1]{\textbf{#1}}

\author{Finn Lindgren \\ University of Edinburgh \\ \texttt{Finn.Lindgren@ed.ac.uk}
   \and Fabian Bachl \\ University of Edinburgh
   \and Janine Illian \\ University of Glasgow
   \and Man Ho Suen \\ University of Edinburgh
   \and H{\aa}vard Rue \\ King Abdullah University of Science and Technology
   \and Andrew E. Seaton \\ University of Glasgow}

\title{\pkg{inlabru}: software for fitting latent Gaussian models with non-linear predictors}

\newcommand{\bm}[1]{\boldsymbol{#1}}
\newcommand{\wt}[1]{\widetilde{#1}}
\newcommand{\ol}[1]{\overline{#1}}
\newcommand{\wh}[1]{\widehat{#1}}
\DeclareMathOperator*{\argmax}{arg\,max}
\DeclareMathOperator*{\tr}{tr}
\newcommand{\proper}[1]{\mathsf{#1}}
\newcommand{\pExp}{\proper{Exp}}
\newcommand{\pN}{\proper{N}}
\newcommand{\pPo}{\proper{Po}}
\newcommand{\pGa}{\proper{Ga}}
\newcommand{\pE}{\proper{E}}

\newcommand{\pCov}{\mathsf{Cov}}
\newcommand{\pKL}[2]{\mathsf{D}_\mathsf{KL}\left(#1\,\middle\|\,#2\right)}

\newcommand{\mmd}{\mathrm{d}}
\newcommand{\md}{\,\mmd{}}

\newtheorem{theorem}{Theorem}

\newcommand{\pUnif}{\proper{Unif}}
\usepackage{hyperref}
\IfFileExists{upquote.sty}{\usepackage{upquote}}{}
\begin{document}

\maketitle

\begin{abstract}
The integrated nested Laplace approximation (INLA) method has become a popular approach for computationally efficient approximate Bayesian computation.  In particular, by leveraging sparsity in random effect precision matrices, INLA is commonly used in spatial and spatio-temporal applications.  However, the speed of INLA comes at the cost of restricting the user to the family of latent Gaussian models and the likelihoods currently implemented in \pkg{INLA}, the main software implementation of the INLA methodology.

\pkg{inlabru} is a software package that extends the types of models that can be fitted using INLA by allowing the latent predictor to be non-linear in its parameters, moving beyond the additive linear predictor framework to allow more complex functional relationships.  For inference it uses an approximate iterative method based on the first-order Taylor expansion of the non-linear predictor, fitting the model using INLA for each linearised model configuration.

\pkg{inlabru} automates much of the workflow required to fit models using \pkg{R-INLA}, simplifying the process for users to specify, fit and predict from models.  There is additional support for fitting joint likelihood models by building each likelihood individually. \pkg{inlabru} also supports the direct use of spatial data structures, such as those implemented in the \pkg{sf} and \pkg{terra} packages.

In this paper we outline the statistical theory, model structure and basic syntax required for users to understand and develop their own models using \pkg{inlabru}.  We evaluate the approximate inference method using a Bayesian method checking approach.  We provide three examples modelling simulated spatial data that demonstrate the benefits of the additional flexibility provided by \pkg{inlabru}.
\end{abstract}


\section{Introduction}

The approximate Bayesian inference package \pkg{INLA} \citep{rue_ApproximateBayesianInference_2009} is an implementation of the integrated nested Laplace approximations (INLA) method of inference for latent Gaussian models (LGMs).  INLA is a fast and accurate approximate inference method that provides an alternative to MCMC for Bayesian inference on LGMs. INLA uses computational methods for sparse Gaussian Markov random fields (GMRFs) \citep{rue_GaussianMarkovRandom_2005} that greatly speed up inference for models with large numbers of latent Gaussian parameters. This efficiency has led to INLA being a go-to choice over MCMC in many contexts since, for the class of LGMs, the computational benefits of INLA far outweigh the minimal approximation error. As such, INLA is a popular choice in spatio-temporal modelling \citep{lindgren_SPDEApproachGaussian_2022, krainski_AdvancedSpatialModeling_2018, bakka_SpatialModelingRINLA_2018, blangiardoSpatialSpatiotemporalModels2013} and has been applied in a diverse set of fields such as ecology \citep{martino_IntegrationPresenceonlyData_2021, illianFittingComplexEcological2013}, astronomy \citep{levis_InferenceBlackHole_2021}, public health \citep{moraga_GeospatialHealthData_2019}, conditional extremes \citep{simpson_HighdimensionalModelingSpatial_2023}, seismology \citep{naylor_BayesianModelingTemporal_2023}, econometrics \citep{bivand_ApproximateBayesianInference_2014} and more.

However, while the class of LGMs is large, as evidenced by the wide variety of applications, there are many models that cannot be fitted using the INLA methodology.  \pkg{inlabru} is a software package that extends the class of models that can be fitted using INLA by allowing the predictor to be a non-linear (deterministic) function of the latent Gaussian parameters \citep{bachl2019inlabru}.  Inference on such models is achieved by an iterative fitting scheme by using INLA to fit successive linearised model configurations.

This development opens up many new modelling possibilities for fitting models with INLA.  The additional flexibility of non-linear predictors allows users to parameterise relationships between predictors and response variables in a bespoke and context-informed way.  Traditionally, software for fitting LGMs requires the predictor to be linear in its parameters (e.g.\ \pkg{mgcv} \citep{Wood_GAMS_2017}, \pkg{glmmmTMB} \citep{glmmTMB_2017}, and \pkg{lme4} \citep{lme4_2015}. However, non-linear relationships appear in many applications, such as modelling the detectability of animals in ecological field surveys \citep{martino_IntegrationPresenceonlyData_2021, yuan_PointProcessModels_2017, bucklandDistanceSamplingMethods2015, millar_EstimatingSizeselectionCurves_1999}, exposure-response models in public health modelling \citep{nasari_ClassNonlinearExposureresponse_2016, gasparrini_ModelingExposureLag_2014, ritz_UnifiedApproachDose_2010}, functional response ecology \citep{matthiopoulos_SpecieshabitatAssociationsSpatial_2020, rosenbaum_FittingFunctionalResponses_2018, smout_FunctionalResponseGeneralist_2010, real_EcologicalDeterminantsFunctional_1979} and consumer choice models in economics \citep{feng_ConsumerChoiceModels_2022}.

Non-linear predictors also appear in models such as self-exciting point processes \citep{serafini_ApproximationBayesianHawkes_2023, hawkes_SpectraSelfexcitingMutually_1971}, population dynamics \citep{newman_ModellingPopulationDynamics_2014}, kernel-smoothed effects \citep{bowman1997applied} and lagged-effects models \citep{shumway_TimeSeriesAnalysis_2017}.   \pkg{inlabru} allows parametric relationships such as these to be incorporated into LGMs and estimated efficiently using the INLA methodology.  This opens up the computational efficiency of INLA to a much broader class of applications.

\pkg{inlabru} is a software interface for \pkg{INLA} that greatly simplifies the process of specifying, fitting and sampling from \pkg{INLA} models.  The process of `stack building', familiar to users of \pkg{INLA}, is now entirely automated based on the model definition. \pkg{inlabru} also supports the use of spatial data objects directly when defining, fitting, and sampling from models. Objects of \pkg{sf} \citep{pebesma_simple_2018} and \pkg{terra} \citep{terra_2024} types are supported,
as well as legacy support for \pkg{sp} \citep{pebesma_spatial_2005} and \pkg{raster}
\citep{raster_2023} types.  \pkg{inlabru} also provides support for working with joint likelihood models by separating the process of constructing likelihoods and fitting models.  By providing \code{predict()} and \code{generate()} methods for \pkg{inlabru} models, \pkg{inlabru} greatly simplifies the process of generating predictions from fitted models.

This additional functionality provided by \pkg{inlabru} can be used in combination with several add-on packages that extend the basic \pkg{INLA} latent models, using hooks that tells \pkg{inlabru} how to map between covariate inputs and latent models,
such as \pkg{MetricGraph} for defining and fitting random fields on networks \citep{bolin_metric_2024, MetricGraph2023},
\pkg{dirinla} for Dirichlet regression \citep{dirinla2022}, \pkg{ETAS.inlabru} for fitting self-exciting point process models in seismology \citep{naylor_BayesianModelingTemporal_2023}, and the \pkg{rSPDE} and \pkg{INLAspacetime} packages \citep{rSPDE2023, bolin_CovarianceBasedRational_2024,Krainski_INLAspacetime_2023} that fractional prameter spatial models and non-separable covariance models.

\pkg{inlabru} is therefore both an extension of, and wrapper for, the existing \pkg{INLA} implementation.  The aims of this software paper are to introduce the iterative INLA method for fitting models with non-linear predictors, to motivate the software syntax that simplifies the INLA workflow, and to provide numerous examples to demonstrate what can be achieved using \pkg{inlabru} that users can use as a starting point for their own work.

\subsection{A simple example}
\label{sec-intro-ex}

The following example demonstrates how to define a model, fit it, and generate posterior samples using \pkg{inlabru}.  It contains all the basic building blocks of the \pkg{inlabru} workflow that will be present in later examples.

We use a simulated dataset with observations
\begin{equation}
z_i = \beta + f(\bm{s}_i) + \epsilon_i  ,\quad i=\dots,n,
\end{equation}
where $\bm{s}_i$ is the location of observation $i$, $\beta$ is an intercept parameter, $f$ is a Gaussian random field with Mat\'ern covariance and $\epsilon_i$ is unstructured Gaussian noise.

\begin{figure}[!htb]
	\centering
	\includegraphics{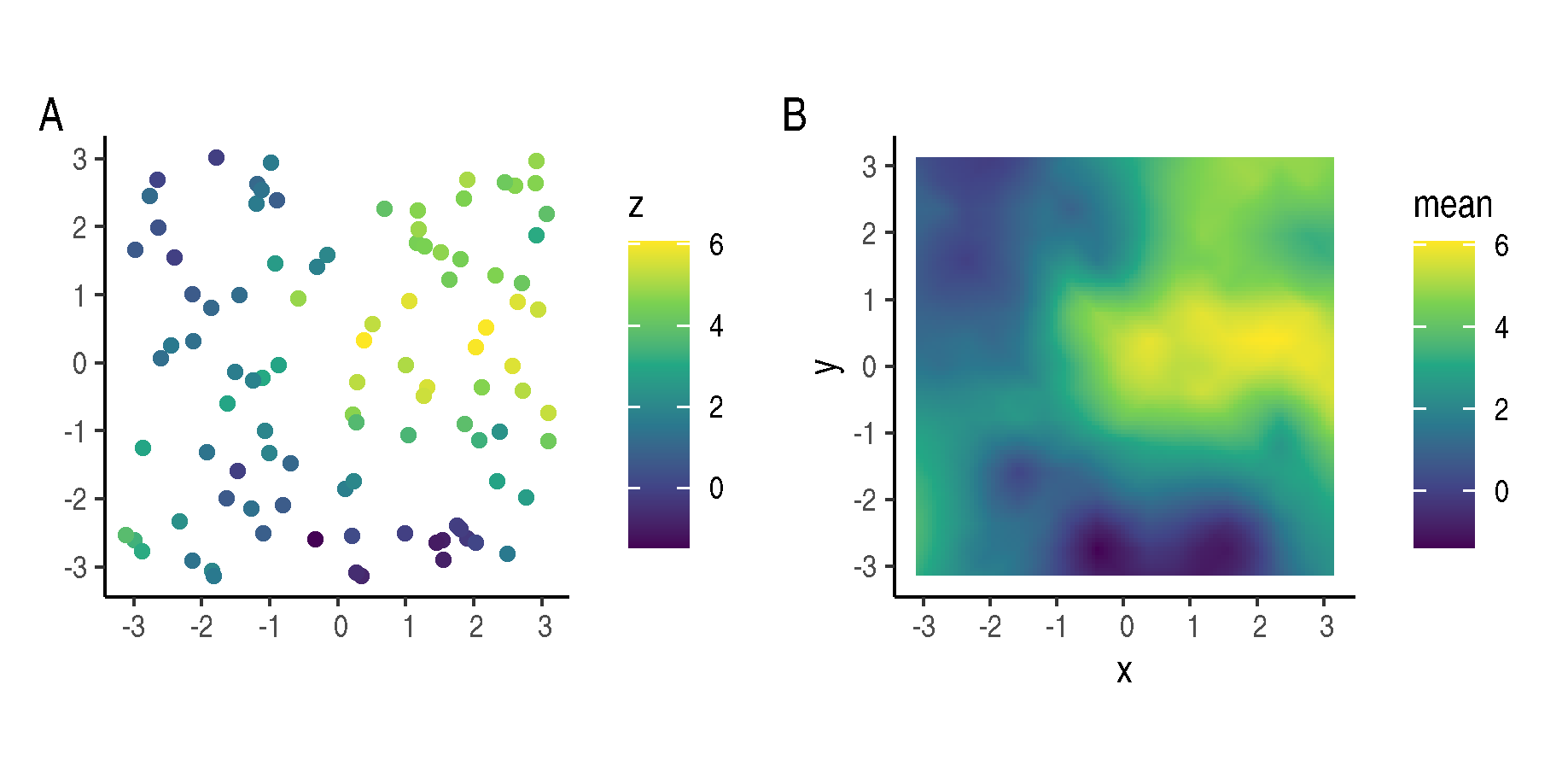}
	\caption{\textbf{A}: simulated point data; \textbf{B}: posterior mean prediction}
	\label{fig-intro-example}
\end{figure}

The data is stored as an \pkg{sf} spatial points object (Figure~\ref{fig-intro-example} A) which can be used directly with \pkg{inlabru}.  The data is included with \pkg{inlabru} under the name \code{toypoints}.  The following code shows how to load the data, fit a model, and generate model predictions.

\begin{Schunk}
\begin{Sinput}
R> library(inlabru)
R> library(INLA)
R> library(sf)
R> 
R> # Load data
R> point_data <- toypoints$points # sf point data
R> mesh <- toypoints$mesh # SPDE mesh
R> pred_locs <- toypoints$pred_locs # sf prediction point locations
R> 
R> # Define model components
R> 
R> # Spatially structure random effect
R> matern <- inla.spde2.pcmatern(mesh,
+    prior.range = c(2, 0.05),
+    prior.sigma = c(2, 0.05)
+  )
R> 
R> cmp <- ~ Intercept(1) +
+    grf(
+      main = geometry,
+      model = matern
+    )
R> 
R> # Construct likelihood
R> lik <- like(
+    formula = z ~ .,
+    data = point_data,
+    family = "gaussian"
+  )
R> 
R> # Fit model
R> fit <- bru(lik,
+    components = cmp
+  )
R> 
R> # Make predictions
R> predictions <- predict(fit,
+    pred_locs,
+    formula = ~ Intercept + grf
+  )
\end{Sinput}
\end{Schunk}

This example contains the essential elements of the \pkg{inlabru} workflow.  The user defines model components using a formula syntax that allows users to choose meaningful names for model components.  Here the labels \code{Intercept} and \code{grf} (an abbreviation of Gaussian random field) were used to name the components of the model but they equally well could be something else.  These are labels for components, not functions, despite the similarity to the function syntax.

The \code{main} argument describes the \textit{input data} for each component.  The spatial information is stored in the \code{geometry} column of the \pkg{sf} object and so this is used when defining the random field component.  The software automatically detects the type of input, the type of component, and interally constructs the prior precision and model design matrix for that component.

An intercept parameter is associated with a column of ones in the model design matrix, and the shorthand notation \code{Intercept(1)} reflects this.  Alternative methods to define an intercept are \code{Intercept(ones, model = 'linear')} if \code{ones} is the name of a covariate in the data or \code{Intercept(rep(1,n), model = 'linear')} if \code{n} is an object stored in the global R environment.  The input data for a component is allowed to be a general R expression song as it returns an object that is appropriate for that type of model component.

The likelihood is constructed in a separate step to model fitting with the likelihood object created by the \code{like()} function.  The \code{sf} point data object is used directly, requiring no pre-processing of data to a standard \code{data.frame}.  This automates the process of `stack building' familiar to \pkg{INLA} users.

After fitting the model, predictions are generated using a \code{predict()} method.  The formula object for the prediction can be a generic R expression that references model components using the user-defined names.  This greatly simplifies the process for generating predictions from INLA models.  Figure~\ref{fig-intro-example} shows the data and the posterior mean prediction.

Figure~\ref{fig-intro-example}B was produced using the \code{predictions} object.  The \code{predict()} method returns an object in the same data format as was used in the predict call which, in this case, is an \code{sf} points object.  Support for plotting \code{sf} data objects is available in the \pkg{ggplot2} package \citep{Wickham_ggplot2_2009}.  The code to produce Figure~\ref{fig-intro-example}B is as simple as
\begin{Schunk}
\begin{Sinput}
R> ggplot() +
+    gg(
+      data = predictions,
+      aes(fill = mean),
+      geom = "tile"
+    ) +
+    scale_colour_viridis_c() +
+    theme_classic()
\end{Sinput}
\end{Schunk}
where \code{mean} is one of the summary statistics returned by \code{predict()}.  See \code{?predict.bru} for a list of all summary statistics.  In general, the prediction data can be in any format for which the component definitions make sense.  This gives users flexibility to conceptually separate data that is used for model fitting from data that is used for prediction.

This example is not simple.  It involves spatial data, a spatially structured random effect, and generating model predictions.  Users already familiar with such models in \pkg{INLA} will recognise that the \pkg{inlabru} code required to run this is more concise and readable.  This allows users to focus more on the model structure rather than wrangling data and design matrices into a format expected by \pkg{INLA}.

The rest of the paper is structured as follows: Section~\ref{INLA-and-iINLA} reviews the standard INLA approach and describes the iterative INLA method for estimation with non-linear predictors, Section~\ref{syntax} covers basic package structure and syntax for specifying, fitting and predicting from models, and Section~\ref{method-checking} presents a Bayesian method checking simulation to validate the iterative INLA inference method.  Section~\ref{examples} presents syntax examples for fitting spatial models.  Examples include a standard LGM with a Besag-York-Mollié random effect, a non-linear predictor example, aggregating a continuously defined random field to discrete areal units, and a joint likelihood model. Finally, Section~\ref{discussion} draws links to other methods that extend the class of models that can be fitted using \pkg{INLA}, discusses existing published research that uses the \pkg{inlabru} non-linear predictors feature, existing software that depends on \pkg{inlabru}, and, finally, ideas for the future development of \pkg{inlabru}.


\section{Iterative INLA}
\label{INLA-and-iINLA}

\subsection{Standard INLA}
\label{standard-INLA}

The INLA method is an approach to compute fast approximate posterior distributions
for Bayesian generalised additive models. The class of models that INLA was developed to address are known as \textit{latent Gaussian models} (LGMs).

The hierarchical structure of a LGM with latent Gaussian vector $\bm{u}$, covariance parameters $\bm{\theta}$, and response variable $\bm{y}$, can be written as
$$
\begin{aligned}
\bm{\theta} &\sim p(\bm{\theta}) \\
\bm{u}|\bm{\theta} &\sim \mathcal{N}\!\left(\bm{\mu}_u, \bm{Q}(\bm{\theta})^{-1}\right) \\
\bm{\eta}(\bm{u}) &= \bm{A}\bm{u} \\
\bm{y}|\bm{u},\bm{\theta} & \sim p(\bm{y}|\bm{\eta}(\bm{u}),\bm{\theta}).
\end{aligned}
$$
The latent Gaussian vector $\bm{u}$ has a covariance structure that depends on hyper parameters $\bm{\theta}$ which have prior distribution $p(\bm{\theta})$. Each linear predictor element, $\eta_i(\bm{u})$, is linked to the latent field via a linear map defined by the known, non-random matrix $\bm{A}$, i.e. the additive predictor for observation $y_i$ is $\eta_i(\bm{u}) = \bm{A}_i \bm{u}$, where $\bm{A}_i$ is the $i$-th row of $\bm{A}$.  This linear predictor defines the location parameter for the distribution of the data $\bm{y}$, via a (possibly non-linear) link function $g^{-1}(\cdot)$. Observations are assumed to be conditionally independent given $\bm{\eta}$
and $\bm{\theta}$, so that $p(\bm{y}|\bm{\eta}(\bm{u}), \bm{\theta}) = \prod_i p(y_i|\eta_i(\bm{u}), \bm{\theta})$.

In the classic formulation of INLA \citep{rue_ApproximateBayesianInference_2009}, the linear predictors are included as part of an augmented latent Gaussian vector $\left[\bm{\eta}, \bm{u}\right]^\intercal$ so that inference on the predictors is included as part of the overall estimation of the latent field.  Since $\bm{\eta}$ is a linear combination of $\bm{u}$, a small amount of noise is added to avoid a singular precision.  The \pkg{INLA} package has recently implemented an alternative approach which avoids the need to constructed the augmented latent Gaussian vector, which can be prohibitively large for certain classes of models \citep{vanniekerk_NewAvenueBayesian_2023, niekerk_LowrankVariationalBayes_2024}.  We present the classic approach below, see Section~\ref{variational-inla} for a brief summary of the recent INLA developments.

INLA is a method to approximate the joint posterior density $p(\bm{\theta} | \bm{y})$, the marginal posterior distributions $(u_j | \bm{y})$ and allows for sampling from the joint posterior $(\bm{u} | \theta)$, where $\bm{u}$ now represents the augmented latent Gaussian vector.
\cite{rue_ApproximateBayesianInference_2009} introduced the approach and \cite{martins_BayesianComputingINLA_2013} includes additional details.

\subsection{The classic INLA method}

The joint posterior density of the hyperparameters is approximated as
$$
\hat{p}(\bm{\theta} | \bm{y}) \propto \left.
\frac{ p(\bm{u}, \bm{\theta}, \bm{y})}{p_G(\bm{u}| \bm{\theta}, \bm{y})} \right|_{\bm{u} = \bm{u}^*(\bm{\theta})},
$$
where $p_G(\bm{u} | \bm{\theta}, \bm{y})$ is the Gaussian approximation to $p(\bm{u} | \bm{\theta}, \bm{y})$ centred at the mode $\bm{u}^*(\bm{\theta})$ for a given $\bm{\theta}$.  This is equivalent to the Laplace approximation of the marginal posterior \citep{tierney_AccurateApproximationsPosterior_1986}.  The marginal posterior $p(\theta_k | \bm{y})$ is approximated using an interpolation of $\hat{p}(\bm{\theta} | \bm{y})$ based on a grid search of the parameter space for $\bm{\theta}$.  For this reason INLA is generally applied in contexts where $\bm{\theta}$ is low dimensional.

The marginal posterior density $p(u_i | \bm{y})$ is approximated as
$$
\hat{p}(u_i | \bm{y}) = \sum_m \hat{p}(u_i | \bm{\theta}^{(m)}, \bm{y}) \hat{p}(\bm{\theta}^{(m)} | \bm{y}) \Delta\bm{\theta}^{(m)},
$$
where $\hat{p}(u_i | \bm{\theta}, \bm{y})$ is an approximate density given below and $\Delta \bm{\theta}^{(m)}$ is an integration weight associated with each grid search location for $\bm{\theta}$, indexed by $m$.

\cite{rue_ApproximateBayesianInference_2007} originally proposed a Gaussian approximation for $\hat{p}(u_i | \bm{\theta}, \bm{y})$ but found that this was not sufficiently accurate.  Instead, \cite{rue_ApproximateBayesianInference_2009} improved on this by using another application of the Laplace approximation:
$$
\hat{p}(u_i | \bm{\theta}, \bm{y}) \propto \left. \frac{ p(\bm{u}, \bm{\theta}, \bm{y})}{p_G(\bm{u}_{-i}| u_i, \bm{\theta}, \bm{y})} \right|_{\bm{u}_{-i} = \bm{u}_{-i}^*(u_i, \bm{\theta})},
$$
where $p_G(\bm{u}_{-i}| u_i, \bm{\theta}, \bm{y})$ is the Gaussian approximation to $p(\bm{u}_{-i}| u_i, \bm{\theta}, \bm{y})$ centred at the mode $\bm{u}_{-i}^*(u_i, \bm{\theta})$, given $u_i$ and $\bm{\theta}$. Until 2022, this was the standard method used by the \pkg{INLA} package.
Although the joint posterior $p(\bm{u} | \bm{y})$ is not available directly, \cite{chiuchiolo_JointPosteriorInference_2023} propose a method generate samples from the joint posterior via a skew-Gaussian copula approach.

\subsection{A new approach for INLA: the variational Bayes correction}
\label{variational-inla}

An alternative to the classic INLA approach was proposed by \cite{vanniekerk_NewAvenueBayesian_2023} and \cite{niekerk_LowrankVariationalBayes_2024} that is more computationally efficient and avoids the need to include $\bm{\eta}$ as part of the latent field.  In this case, the joint posterior is represented as $p(\bm{u},\bm{\theta} | \bm{y}) \propto p(\bm{\theta})p(\bm{u} | \bm{\theta})\prod_i p(y_i | (\bm{A}\bm{u})_i, \bm{\theta})$ and inference for $\bm{\eta}$ is dealt with separately to inference for $\bm{u}$ and $\bm{\theta}$.  In this new approach, \cite{vanniekerk_NewAvenueBayesian_2023} return to the idea of using a Gaussian approximation for $p(u_i | \bm{\theta}, \bm{y})$.  To correct for the inaccuracies identified in \cite{rue_ApproximateBayesianInference_2007}, \cite{niekerk_LowrankVariationalBayes_2024} use a variational Bayes correction to the mean of the Gaussian approximation.  As shown by \cite{vanniekerk_NewAvenueBayesian_2023}, this approach is more computationally efficient and it was made the default mode of inference from version 22.11.22 of \pkg{INLA}.

For a fixed $\bm{\theta}$, the Gaussian approximation to $p(\bm{u} | \bm{\theta}, \bm{y})$ is calculated from a second order Taylor expansion of the likelihood around the mode, with associated precision $\bm{Q}^*(\bm{\theta})$, which has a sparsity structure that depends on the sparsity of the prior precision $\bm{Q}(\bm{\theta})$ as well as the non-zero entries in $\bm{A}^\intercal \bm{A}$.  The Gaussian approximation is then used to approximate the marginal posteriors as $p(u_j | \bm{\theta}, \bm{y}) \approx \pN(\mu(\bm{\theta})_j, \bm{Q}^*(\bm{\theta})^{-1}_{jj})$, where $\bm{\mu}(\bm{\theta})$ is the joint mode of $p(\bm{u} | \bm{\theta}, \bm{y})$.  The posterior density $p(u_j | \bm{y})$ is then approximated by numerical integration over $\bm{\theta}$, just as in the classic formulation of INLA.  For inference on the predictors, the posterior $p(\bm{\eta} | \bm{\theta}, \bm{y})$ is approximated as a Gaussian density with expected value $\bm{A}\bm{\mu}(\bm{\theta})$ and covariance $\bm{A}\bm{Q}^*(\bm{\theta})^{-1} \bm{A}^\intercal$.

Note that although $\bm{Q}^*(\bm{\theta})$ is sparse, its inverse is not.  However, \cite{niekerk_LowrankVariationalBayes_2024} present a computationally efficient method to compute the $j$-th diagonal element of $\bm{A}\bm{Q}^*(\bm{\theta})^{-1} \bm{A}^\intercal$, which is all that is required for the approximation to $p(\eta_j | \bm{\theta}, \bm{y})$ and avoids the need to compute the full inverse. The posterior density $p(\eta_j | \bm{y})$ is then approximated using numerical integration as above.

It is important to note that, although the approximate posteriors densities $p(u_j | \bm{\theta}, \bm{y})$ and $p(\eta_j | \bm{\theta}, \bm{y})$ are Gaussian, the full marginals can deviate from normality due to the numerical integration.  This allows for, say, non-symmetric posteriors to be estimated, even when using these Gaussian approximations.

\subsection{Linear predictors in LGMs}
\label{sec-LGM-predictors}

The structure of LGMs facilitates a modular view of the additive predictor $\bm{\eta}(\bm{u}) = \bm{A}\bm{u}$, which can be decomposed by thinking of each fixed and random effect as individual model components.
By a model component we mean anything that would usually be viewed as a separate additive component of the linear predictor, such as an intercept parameter, the linear effect of a known covariate or a `random effect' component, such as the sparse Gaussian Markov Random Field (GMRF) effects \citep{rue_GaussianMarkovRandom_2005} that are implemented in INLA, examples of which include the Stochastic Partial Differential Equation (SPDE) approach to a GRF with Mat\'ern covariance \citep{lindgren_ExplicitLinkGaussian_2011}, and many other options including conditional auto-regressive and random walk models.

The matrix $\bm{A}$ is the model design matrix.  In presentations of mixed effect models this is often decomposed as $\bm{A} = \left[\bm{X} \bm{Z} \right]$, where $\bm{X}$ contains the covariate information for fixed effects and $\bm{Z}$ contains the information to map the random effects to the observations.  In the Bayesian LGM setting the distinction between fixed and random effects is somewhat arbitrary.  We can think of the design matrix as decomposed into separate sub-matrices for each model component, i.e.\
$$
\bm{A} = \begin{bmatrix} \bm{A}^{(1)} & \dots & \bm{A}^{(d)} \end{bmatrix}
$$
for a model with $d$ components, viewing each model component as having it's own `component design matrix', denoted here as $\bm{A}^{(j)}$ for $j = 1, \dots, d$.
The parameter vector can be partitioned into $\bm{u} = [\bm{u}^{(1)}, \dots, \bm{u}^{(d)}]$, where each $\bm{u}^{(j)}$ is the parameter vector for component $j$.  Given this decomposition the predictor expression for the $i$-th observation can be written as $\eta(\bm{u})_i = \delta_i + \sum_{j=1}^{d} \bm{A}^{(j)}_i \bm{u}^{(j)}$, where $\bm{A}^{(j)}_i$ is the $i$-th row of $\bm{A}^{(j)}$, and $\delta_i$ is an optional constant offset term.

For example, the component design matrix for an intercept parameter is a column vector of ones and the component design matrix for the effect of a known covariate is a column vector containing the covariate information for each observation.
The component design matrix for an iid Gaussian effect is the identity matrix (or a block diagonal matrix consisting of identity sub-matrices of the correct dimension for each group).
The component design matrix for an SPDE effect contains the finite element basis functions evaluated at each observation location.
The key idea is that each component of the model predictor can be written as a known matrix multiplied by a parameter vector.

This view is the basis for the implementation in \pkg{inlabru} that automatically constructs the full model design matrix, given the component definitions and data.
Each type of component has an associated method for constructing the component design matrix.
These are known as \code{bru\_mapper()} methods as they `map parameters to observations'.
The user only needs to provide a model component definition and the data and the software does the rest.
This component definition can be conceptualised in the mind of the user as `the linear effect of covariate x' or `an SPDE effect on latitude and longitude', for example, with an associated syntax for writing these as an \proglang{R} \code{formula} object.
The software then parses these definitions and applies the relevant \code{bru\_mapper()} methods to the data to construct the appropriate component design matrices and full model design matrix.
In addition to linear effects, \pkg{inlabru} also supports non-linear component effects, e.g.\
via marginal distribution mappers, that are discussed in Supplement~\ref{sec:supplement:mapper-intro}.

In the existing implementation of \pkg{INLA}, for all but the most simple models, users must construct the design matrix with the use of helper functions, in a process that is known as `stack building'
\citep[Section~2.5]{lindgren_BayesianSpatialModelling_2015}.  For complex models this process can be quite involved.
A key feature of \pkg{inlabru} is that this is now entirely automated, greatly simplifying the model fitting process.

For certain model components, such as the SPDE effect, the design matrix has become known among INLA users as the `projector' matrix, as it is thought of as `projecting the GMRF parameters to the observation locations'.
In this paper we use the term projector matrix and design matrix interchangeably.

\subsection{Approximate INLA for non-linear predictors}
\label{iterative-INLA}

\pkg{inlabru} extends the class of models that can be fitted using INLA by allowing the predictor to be a non-linear function of the parameter vector $\bm{u}$.  The premise for the \pkg{inlabru} implementation for non-linear predictors is to build on the existing \pkg{INLA} implementation using an iterative model fitting scheme, fitting the model using \pkg{INLA} at each iteration.  \pkg{inlabru} can therefore be viewed as both a wrapper for, and an extension of, the existing \pkg{INLA} implementation. This iterative approach is an approximate method of inference based on a linearisation step and the properties of the resulting approximation will depend on the nature of the non-linearity and the choice of convergence criteria.

Let $\wt{\bm{\eta}}(\bm{u})$ denote a non-linear predictor, i.e.\ a deterministic function of $\bm{u}$.  Choosing some linearisation point $\bm{u}_0$, the 1st order Taylor approximation of $\wt{\bm{\eta}}(\bm{u})$ at $\bm{u}_0$ can be written as
\begin{align*}
\ol{\bm{\eta}}(\bm{u})
&=
\wt{\bm{\eta}}(\bm{u}_0) + \sum_{j=1}^d \bm{B}^{(j)}(\bm{u}^{(j)} - \bm{u}_0^{(j)})
=
\left[\wt{\bm{\eta}}(\bm{u}_0) - \sum_j \bm{B}^{(j)}\bm{u}_0^{(j)}\right] +\sum_j \bm{B}^{(j)} \bm{u}^{(j)}
\\&=
\bm{\delta} + \sum_j \bm{B}^{(j)} \bm{u}^{(j)}
,
\end{align*}
where $\bm{B}^{(j)}$ are the derivative matrices of the non-linear predictor,
with respect to each latent component state vector $\bm{u}^{(j)}$, evaluated
at the linearisation point $\bm{u}_0$. In contrast to the linear predictor case,
the offset term $\bm{\delta}$ is not optional, as it varies with the linearisation
point and is therefore typically non-zero.

Generally we define the observation models via
\begin{align*}
\bm{y} | \bm{u}, {\bm{\theta}} &\sim p(\bm{y} | \wt{\bm{\eta}}(\bm{u}), {\bm{\theta}}),
\end{align*}
so that the observation distribution only depends on the latent vector $\bm{u}$
via the predictor $\wt{\eta}(\bm{u})$.  Commonly, a link function transformation
is applied to each element of the predictor vector, so that observation $y_i$ is
linked to $g^{-1}\left[\wt{\eta}(\bm{u})_i\right]$. However, for some models, such as inhomogeneous Poisson point processes, further linear or non-linear transformations may also be applied internally, in order to setup the required likelihood construction in for INLA \citep{yuan_PointProcessModels_2017}.

The non-linear observation model
$p(\bm{y}|\wt{\bm{\eta}}(\bm{u}),\bm{\theta})$
is approximated by replacing the non-linear predictor with its linearisation,
so that the linearised model is defined by
$$
\ol{p}(\bm{y}|\bm{u},\bm{\theta})
=
p(\bm{y}|\ol{\bm{\eta}}(\bm{u}),\bm{\theta})
\approx
p(\bm{y}|\wt{\bm{\eta}}(\bm{u}),\bm{\theta})
=
\wt{p}(\bm{y}|\bm{u},\bm{\theta}).
$$

Each such linearised predictor model $\ol{\eta}(\bm{u})$ has design matrix
\begin{align*}
\bm{B} &= \begin{bmatrix} \bm{B}^{(1)} & \dots & \bm{B}^{(d)} \end{bmatrix}
\end{align*}
and offset term $\bm{\delta}=\wt{\bm{\eta}}(\bm{u}_0) - \bm{B}\bm{u}_0$.
This defines a model that can be fitted using the existing INLA methodology for LGMs with linear predictors.
The next step is to choose a suitable linearisation point. This is done using a fixed point iteration method,
which locates a point $\bm{u}_0$ such that the resulting conditional posterior mode $\wh{\bm{u}}$
of the linearised model is the same as the linearisation point. Starting at some initial point $\bm{u}_0^{(0)}$, for $k=1,2,\dots$ applying the INLA method, to obtain
\begin{align*}
    \wh{{\bm{\theta}}}^{(k)} &= \argmax_{\bm{\theta}} \ol{p}( {\bm{\theta}} | \bm{y} ; \bm{u}_0^{(k-1)}), \\
    \wh{\bm{u}}^{(k)} &= \argmax_{\bm{u}} \ol{p}(\bm{u} | \bm{y}, \wh{{\bm{\theta}}}, \bm{u}_0^{(k-1)}),
\end{align*}
and choosing the next linearisation point as the point $\bm{u}_0^{(k)} = (1-\alpha) \bm{u}_0^{(k-1)}+\alpha\wh{\bm{u}}^{(k)}$ for some $\alpha>0$ that minimises a norm of the difference between the linearised and non-linear predictors. The iteration continues until convergence is detected. The convergence stopping criteria
are when
\begin{itemize}
\item a maximum number of iterations have been carried out (the \code{bru\_max\_iter} option, default $10$), or
\item the relative maximum change in the linearisation point is below a threshold relative to the estimated posterior standard deviations (\code{bru\_method\$rel\_tol}, default 10\%), and
\item the line search for choosing $\alpha$ is \emph{inactive}, i.e.\ $\alpha\approx 1$, indicating that the non-linear and linearised predictors coincide.
\end{itemize}
For more details on the iterative method, see Supplementary Material Section~\ref*{sec:supplement:iteration}.

\subsection{Approximation accuracy}
\label{sec-approximation-accuracy}

Whereas the inlabru optimisation method leads to an estimate where $\| \wt{\bm{\eta}} (\bm{u}_*) - \ol{\bm{\eta}}(\bm{u}_*)\|=0$ for a specific $\bm{u}_*$, the overall posterior approximation accuracy depends on the degree of nonlinearity in the vicinity of $\bm{u}_*$.
There are two main options for evaluating this nonlinearity, that can both be computed using sampling from the approximate posterior distribution.
The first option is
$$
\begin{aligned}
\sum_i \frac{E_{\bm{u}\sim \ol{p}(\bm{u}|\bm{y})}\left[
|\ol{\bm{\eta}}_i(\bm{u})-\wt{\bm{\eta}}_i(\bm{u})|^2\right]}{\mathrm{Var}_{\bm{u}\sim \ol{p}(\bm{u}|\bm{y})}(\ol{\bm{\eta}}_i(\bm{u}))} ,
\end{aligned}
$$
which is the posterior expectation of the component-wise variance-normalised squared deviation between the non-linear and linearised predictor. Note that the normalising variance includes the variability induced by the posterior uncertainty for $\bm{\theta}$, whereas the $\|\cdot\|_V$ norm used for the line search used only the posterior mode.

The second option is to approximate the Kullback-Leibler divergences between
the conditional posterior distributions under the linear and nonlinear models,
\begin{align}
\label{eq-KL-1}
\pKL{\ol{p}}{\wt{p}} &=
E_{\bm{u}\sim \ol{p}(\bm{u}|\bm{y},\bm{\theta})}
\left[\ln \frac{\ol{p}(\bm{u} |\bm{y},{\bm{\theta}})}{\wt{p}(\bm{u}|\bm{y},{\bm{\theta}})}\right], \\
\label{eq-KL-2}
\pKL{\wt{p}}{\ol{p}} &=
E_{\bm{u}\sim \wt{p}(\bm{u}|\bm{y},\bm{\theta})}
\left[\ln \frac{\wt{p}(\bm{u} |\bm{y},{\bm{\theta}})}{\ol{p}(\bm{u}|\bm{y},{\bm{\theta}})}\right],
\end{align}
both evaluated at the posterior mode for $\bm{\theta}$.
While it may be possible to also approximate the K-L divergence for the full posterior distributions, in practice it is easier to evaluate the conditional K-L divergences only at the posterior mode for $\bm{\theta}$.
Implementing this requires access to specific aspects of
the likelihood and prior distribution details, that are available from the \pkg{INLA} software package output, and we plan to include an implementation of these K-L divergence as a model diagnostic in a future version of \pkg{inlabru}.
Supplementary Material Section~\ref*{sec:supplement:accuracy} presents this in more detail and includes results to approximate Equations~\eqref{eq-KL-1} and~\eqref{eq-KL-2}.

Users can also evaluate the accuracy of the method for a given model using general Bayesian method checking approaches, as we show in Section~\ref{method-checking}.

\subsection{Well-posedness and initialisation}
On a side note, one might be concerned about initialisation at, or convergence to, a saddle point. Although it is not implemented in inlabru, we want to talk about the technicality how we define the initial linearisation point $\bm{u}_0$.

Generally speaking, any values of $\bm{u}_0$ work except the case that the gradient evaluated at $\bm{u}_0$ is $\bm{0}$ because the linearisation point will never move away if the prior mean is also $\bm{0}$. In general, this tends to be a saddle point problem. In some cases the problem can be handled by changing the predictor parameterisation or just changing the initialisation point using the \code{bru\_initial} option. However, for true saddle point problems, it indicates that the predictor parameterisation may lead to a multimodal posterior distribution or is ill-posed in some other way. This is a more fundamental problem that cannot be fixed by changing the initialisation point.

In these examples, where $\beta$ and $\bm{u}$ are latent Gaussian components, the predictors 1, 3, and 4 would typically be safe, but predictor 2 is fundamentally non-identifiable:
$$
\begin{aligned}
\bm{\eta}_1 &= \bm{u}, \\
\bm{\eta}_2 &= \beta \bm{u}, \\
\bm{\eta}_3 &= e^\beta \bm{u}, \\
\bm{\eta}_4 &= F_\beta^{-1} ( \Phi(z_\beta)) \bm{u}, \quad z_{\beta} \sim \mathsf{N}(0,1) .
\end{aligned}
$$


\section{Package structure and basic syntax}
\label{syntax}

The main workhorse of \pkg{inlabru} is the \code{bru()} function, which is used to fit models.  \code{bru()} automatically detects non-linear predictors and, if required, runs the iterated INLA procedure.
The \code{bru()} function requires a \code{components} argument, which is an \proglang{R} formula object specifying the model components.  In addition to the components specification, \code{bru()} has an optional \code{formula} argument that specifies how to combine the model components, possibly in a non-linear way, to construct the latent predictor.
If no \code{formula} is provided the model assumes a linear additive predictor based on the \code{components} formula.  In this case \code{bru()} internally constructs the model design matrix and makes a single call to \code{INLA::inla()} to fit the model.

Below we summarise the basic workflow to specify, fit and sample from models using \pkg{inlabru}.

\subsection{Defining, estimating and sampling from models}

\subsubsection{Define model components}

The syntax for defining model components is directly related to the perspective on LGM predictors presented in Section~\ref{sec-LGM-predictors}.  Each model component, which is a single latent Gaussian parameter, or a collection of related parameters, has an associated prior covariance structure as well as a component design matrix to relate parameters to observations.

Similar to other GAM fitting software, models components are defined via a formula-like syntax.
Each component can be specified by including the following expression in a formula object:
\begin{Schunk}
\begin{Sinput}
R> component_name(
+    main = ...,
+    model = ...,
+    ...
+  )
\end{Sinput}
\end{Schunk}
where \code{component\_name} is a user-chosen name for the component and does not refer to a specific function name.

The \code{main} argument takes an \proglang{R} expression that specifies the \textit{input data} for the component.  For example, \code{main} could be the name of a covariate for a linear fixed effect. Or it could be the name of a covariate that includes indices for a latent random effect, such as a time variable for an auto-regressive model component.  For a spatial effect component it could be the name of the geometry column in an \code{sf} data object or an \proglang{R} expression like \code{cbind(x, y)} where \code{x} and \code{y} are the names of coordinate variables, for effects such as the SPDE where a 2-column matrix of coordinates is expected.  In other words, \code{main} specifies key information that is required to build the component design matrix.

The flexibility of allowing \code{main} to be an \proglang{R} expression is the basis for the support for various spatial data types and allows \pkg{inlabru} to be readily extended to support other data structures in the future. We cover this feature in more detail below. The optional arguments \code{group} and \code{replicate} also accept general \proglang{R} expressions but for the corresponding features of the \code{INLA::f()} function.  This allows users flexibility in specifying the model and reduces the amount of pre-processing of data required compared to \pkg{INLA}.

The information provided by \code{main} is then used by \code{bru\_mapper()} internally to construct the relevant design matrix.  Exactly what form this takes depends on the type of model component which is specified using the \code{model} argument.  Most models are identified using a character string.  For example, \code{model = "linear"}, defines a component that is the linear effect of the data provided in \code{main}.  Similarly, \code{model = "ar1"} defines a auto-regressive order 1 model component.  The full list of supported component names can be found by running \code{INLA::inla.list.models()\$latent}.  An exception to this is certain spatial models defined by the SPDE random effect.  In this case, \code{model} can be a SPDE model object.

The user-chosen name \code{component\_name} appears in model summaries and can be used in prediction formulas, or when sampling from the posterior.  This makes it easy for users to track which latent parameters correspond to which components, as the model object uses this label.  This is in contrast to \pkg{INLA} which does not have the ability for user-defined naming of each component and requires the user to use their own indexing system to track which latent parameters correspond to each component.

We now give some invented examples to clarify the above template for defining model components in specific situations.

Suppose we have covariate named \code{salt} included in the data and we wish to include the linear effect of salt as a component of the model predictor.  Additionally, suppose that the only other parameter in the predictor is an intercept component.  Then the model components would for this model could defined as

\begin{Schunk}
\begin{Sinput}
R> cmps <- ~ effect_of_salt(salt,
+                           model = "linear") +
+    intercept(1, model = "linear")
\end{Sinput}
\end{Schunk}

The name \code{effect\_of\_salt} is used to label relevant parts of the fitted model object, will appear in model summaries, and can be used in the formula passed to the \code{predict()} methods for fitted models (which we cover in more detail in the full examples below).  Note that \pkg{inlabru} interprets \code{intercept(1,...)} as equivalent to \code{intercept(rep(1, n), ...)} where \code{n} is the number of rows in the model design matrix.  This is shorthand to make defining intercept-like components more straightforward.

The model types can be any of the models that are accepted by the \code{INLA::f()} function, as well as the additional \pkg{inlabru} specific options of \code{"constant"}, \code{"offset"} (a backwards compatibility synonym for \code{"constant"}), \code{"linear"}, \code{"factor\_full"}, \code{"factor\_contrast"}.  Additionally the \code{"fixed"} component type allows users to specify fixed effects using standard \code{lm()}-type syntax such as \code{~ x1*x2} for an interaction.

The additional (optional) arguments to component definitions are any that can be accepted by \code{INLA::f()}, such as the already mentioned \code{group} and \code{replicate} arguments, but also prior specifications, linear constraints and initial values, for example.

We can see from the above that multiple model components as additive terms in a formula object that is then passed to the \code{like()} function to construct the model likelihood.  In general, any number of components can be defined by specifying model definitions additively in the components formula object:
\begin{Schunk}
\begin{Sinput}
R> cmp <- ~ component_1(main = ..., model = ..., ...) +
+    component_2(main = ..., model = ..., ...) +
+    ...
\end{Sinput}
\end{Schunk}

There are a number of conventions that simplify the process of defining common model components such as linear fixed effects and intercept-like parameters.

A component defining the linear effect of a covariate can also be defined simply by using the name of the covariate in the formula. e.g. \code{cmp <- ~covariate\_name} is equivalent to \code{cmp <- ~covariate\_name\_effect(main = covariate\_name, model = "linear")}.
This simplifies the approach to specify common model components and is similar to other GAM fitting software in \proglang{R}. However, we prefer to present the more verbose definition as it more closely resembles the definition of more complicated model components.

For a final more complicated example, consider the case where we have a temporal covariate named \code{season} which takes values $1,2,3,4$ which correspond to the four seasons, and a discrete spatial index given by a covariate \code{sp\_index} which takes values $1, 2, \ldots, M$ with associated neighbourhood graph object \code{sp\_graph}.  Then to define a spatio-temporal Besag CAR model with an $\text{AR}(1)$ temporal dependence structure we can include the following expression in the component definitions:

\begin{Schunk}
\begin{Sinput}
R> cmp <- ~ sp_season_comp(sp_index,
+                          model = "besag",
+                          graph = sp_graph,
+                          group = season,
+                          group.model = "ar1"
+  )
\end{Sinput}
\end{Schunk}

Note that this has the same basic structure as the definition of a linear model component above.  There is a main input, and a model definition.  In addition to this, we include a \code{group} variable and a \code{group.model} that define the temporal dependence model, as well as the \code{graph} argument which is expected for the Besag component type.  \pkg{inlabru} automatically parses this expression into a component design matrix when constructing the model likelihood.  Note that this greatly simplifies the usual process of defining such effects in \pkg{INLA} which require users to construct the relevant component design matrices directly and the full model design matrix using \code{inla.stack()}.  All of this is now automated and the user can focus on defining the model.

The documentation for defining model components can be found by running \code{?component}.  Any information that is not included there is likely to be found in \code{INLA::f} which is the equivalent documentation in \pkg{INLA}.

\subsubsection{Define model formula}

The model formula specifies the response variable and how the model components are combined to form the predictor expression.
For example, a standard additive predictor with response variable \code{y} has the form
\begin{Schunk}
\begin{Sinput}
R> fml <- y ~ component_1 + component_2 + ...
\end{Sinput}
\end{Schunk}
Note that each model component is referred to by the user-chosen name specified in the components definition.  In this example the names are just \code{component\_1} and \code{component\_2}. The predictor formula can be any \proglang{R} expression, including non-linear functions of components.
For example, a user could define a product of two latent components by using
\begin{Schunk}
\begin{Sinput}
R> y ~ component_1 * component_2
\end{Sinput}
\end{Schunk}
Alternatively, the user could also define a new function in the environment and use this within the formula expression.  For example, an equivalent to the above product is
\begin{Schunk}
\begin{Sinput}
R> my_prod <- function(a, b) a * b
R> fml <- y ~ my_prod(component_1, component_2)
\end{Sinput}
\end{Schunk}
This allows more complex predictor expressions to be stored as separate functions rather than directly specified in the formula expression. Note that despite similar appearance to other GAM/GLM software conventions, the syntax \code{component\_1 * component\_2} is \underline{not} defining an interaction of two covariates.  Instead this formula object is parsed as a literal function of the model components.  In this case the model has two components, and the predictor expression is the multiplication of them with each other.  Using the component design matrix notation we defined above, this predictor can be written mathematically as $\wt{\bm{\eta}}(\bm{u}) = \bm{A}^{(1)} \bm{u}^{(1)} \odot \bm{A}^{(2)} \bm{u}^{(2)}$, where $\odot$ denotes element-wise multiplication.

In order to calculate the derivatives matrix required for the linearisation step, \pkg{inlabru} uses numerical derivatives rather than requiring the user to directly specify the derivatives of the predictor formula. The component mapper system computes the linearised component offset and design matrices
discussed in Supplement~\ref{sec:supplement:mapper-intro}, followed by numerical differentiation of the combined predictor formula, and subsequent combination using the chain rule.
In future this could be improved by linking \pkg{inlabru} with automatic differentiation libraries or allowing users to specify their own derivative functions directly.  We advise users to define predictor expressions that are at least twice differentiable since this is a requirement of the Laplace approximation.  However this is a minimum requirement and is not a guarantee of good approximation properties for all twice-differentiable non-linear functions. At the moment no checks are made as to the differentiability of the predictor formula and it is up to users to make sure they define differentiable expressions.

\subsubsection{Construct the likelihood and fit the model}

To facilitate models with multiple likelihoods with shared components, \pkg{inlabru} contains a likelihood object class that is constructed using the \code{like()} function.
To fit a model the components definition and likelihood(s) are then passed to the fitting function \code{bru()}.
\begin{Schunk}
\begin{Sinput}
R> lik <- like(
+    formula = fml,
+    data = ...,
+    family = ...,
+    ...
+  )
R> fit <- bru(lik,
+    components = cmp
+  )
\end{Sinput}
\end{Schunk}

The formula is automatically inspected to check if it is non-linear and, if so, \code{bru()} uses the iterated INLA inference approach.
The \code{family} argument specifies the data likelihood.
A full list of currently implemented likelihoods can be viewed by running \code{INLA::inla.list.models("likelihood")}

For models with a single likelihood, an alternative is to call \code{bru()} directly
\begin{Schunk}
\begin{Sinput}
R> fit <- bru(
+    components = cmp,
+    formula = fml,
+    family = ...,
+    data = ...
+  )
\end{Sinput}
\end{Schunk}

However, all the remaining example code in this paper and in the accompanying code repository uses the \code{like()} function to specify models.

Separating out likelihood construction as a separate task to model fitting facilitates support for multiple likelihood models which can be specified as simply as
\begin{Schunk}
\begin{Sinput}
R> lik_1 <- like(
+    formula = fml_1,
+    data = data_1,
+    family = ...
+  )
R> 
R> lik_2 <- like(
+    formula = fml_2,
+    data = data_2,
+    family = ...
+  )
R> 
R> fit <- bru(
+    lik_1, lik_2,
+    components = cmp,
+    formula = fml
+  )
\end{Sinput}
\end{Schunk}
Note that the components are only specified once directly in the call to \code{bru()}.  The likelihoods each have their own predictor formula object that use some or all of these components.
In this way it is possible to specify model components that are shared across multiple likelihoods.

Recall that the input data for each component is define using a general \proglang{R} expression.
This functionality greatly simplifies the specification of joint models since the predictor expressions can be constructed for each likelihood but each component need only be defined once.
The \code{main} argument for each component specifies the means of constructing the required input data and projection matrices when evaluated on either \code{data\_1} or \code{data\_2}.

\subsubsection{Sampling from the posterior}

The \code{generate()} function can be used to sample from the model posterior.  The following function produces 500 samples:
\begin{Schunk}
\begin{Sinput}
R> generate(
+    fit,
+    data = ...,
+    formula = ...,
+    n.samples = 500
+  )
\end{Sinput}
\end{Schunk}
with the parameters sampled depending on what is included in the \code{formula} argument.  The formula can be any \proglang{R} expression that uses named model components, which can be useful for investigating functional summaries of model parameters.  The \code{predict()} method can be used to calculate summary statistics of posterior Monte Carlo samples, such as the empirical mean, mode, standard deviation and quantiles.

The \code{data} argument can be used to specify the input data for the predictions, such as specific spatial locations or covariate values, for example.
To sample directly from latent parameters without specifying any new input data (or, in other words, to sample with an identity projection matrix), the keyword \code{\_latent} can be appended to a latent component name.  For example
\begin{Schunk}
\begin{Sinput}
R> generate(
+    fit,
+    formula = ~ component_name_latent,
+    n.samples = 500
+  )
\end{Sinput}
\end{Schunk}
will generate 500 samples of the latent Gaussian parameters associated with the component named \code{component\_name}.

Similarly, individual components can be evaluated for arbitrary inputs by using the keyword \code{\_eval} appended to a latent component name.
For example,
\begin{Schunk}
\begin{Sinput}
R> generate(fit,
+    formula = ~ component_name_eval(1:5),
+    n.samples = 500
+  )
\end{Sinput}
\end{Schunk}
generates samples of the component named \code{component\_name} evaluated with input data \code{c(1,2,3,4,5)}.
When using this functionality care should be taken to ensure that the input data are in the correct format for the component model type.

\subsection{Support for spatial data}

As mentioned above, \pkg{inlabru} provides support for \pkg{sf} \citep{pebesma_simple_2018} and \pkg{terra} \citep{terra_2024} data structures. This means that \pkg{sf} and \pkg{terra} objects can be passed as data and covariates, respectively, directly to the \pkg{like()} and \pkg{bru()} functions.  For example, a log-Gaussian Cox process can be fitted using an \code{sf} points object to describe the locations of the observed points and an \code{sf} polygon object to define the observation window. Spatial objects can also be used for prediction as data objects in \code{predict()} and \code{generate()} calls.

Some model components can work directly with the spatial information provided in the geometry column of the \code{sf} data object.  For example, the \code{bru\_mapper()} method for the SPDE random effect has methods for working with \code{sf} geometry objects.  The mapper is system is the basis for extending support for spatial data to more model components in the future.

The \code{eval\_spatial()} methods can be used to extract information from spatial data objects.  This is useful for defining components where spatial data is an input.  For example, a component specified as
\begin{Schunk}
\begin{Sinput}
R> ~ my_sp_effect(
+    main = a_spatial_object,
+    model = "linear"
+  )
\end{Sinput}
\end{Schunk}
is equivalent to
\begin{Schunk}
\begin{Sinput}
R> ~ my_sp_effect(
+    main = eval_spatial(a_spatial_object, .data.),
+    model = "linear"
+  )
\end{Sinput}
\end{Schunk}
where \code{.data.} is a keyword that refers to whatever data object is passed to the \code{like()} or \code{bru()} function.  The \code{eval\_spatial()} function extracts the value of the spatial covariate at the locations provided in the data. See \code{?eval\_spatial} for more information on what spatial data formats are supported.

There is also legacy support for the \pkg{sp} and \pkg{raster} packages.  For these legacy packages there is support for the plotting of spatial data.  The \code{gg()} function is a geometry class for \pkg{ggplot2} \citep{Wickham_ggplot2_2009} plot objects.  It automatically detects the type of \pkg{sp} object (points, lines, pixels, grid or polygons) and coerces this into a correct format for plotting with standard \code{ggplot2} geometry classes.  This plotting support is not required for the newer spatial packages which have already been supported in \pkg{ggplot2} with the \code{geom\_sf} function, the \pkg{tidyterra} package \citep{tidyterra}, and the \code{geom\_fm} function in the \pkg{fmesher} package \citep{fmesher_2024}.


\section{Approximate Bayesian method checking}
\label{method-checking}

As a proof of concept test we compare the approximate posterior distributions
to the truth in a simple case where consider the true posterior distribution
has a known form.
Consider the hierarchical model of the form
$$
\begin{aligned}
\lambda &\sim \pExp(\gamma) \\
(y_i|\lambda) &\sim \pPo(\lambda),
\end{aligned}
$$
for $i = 1, \ldots, n$.
Treating $\gamma$ as a known constant, the posterior density is
$$
\begin{aligned}
p(\lambda | \{y_i\}) &\propto p(\lambda, y_1,\dots,y_n) \\
&\propto \exp(-\gamma\lambda) \exp(-n\lambda) \lambda^{n\ol{y}} \\
&= \exp\{-(\gamma+n)\lambda\} \lambda^{n\ol{y}},
\end{aligned}
$$
where $\ol{y}=\frac{1}{n}\sum_{i=1}^n y_i$.  This is proportional to a $\pGa(\alpha = 1+n\ol{y}, \beta = \gamma+n)$ density.

This model can be reparameterised by introducing a latent Gaussian variable $u \sim \pN(0,1)$.  Using the inverse cumulative distribution function (CDF) for the exponential distribution we can rewrite the model as
$$
\begin{aligned}
\label{eq-poisson-exp}
\lambda(u) &=-\ln\{1-\Phi(u)\}/\gamma \\
y_i|u &\sim \pPo(\lambda(u)),
\end{aligned}
$$
where $\Phi(u)$ is the CDF of a standard normal distribution.  Using a log link, the non-linear predictor function is $\wt{\eta}(u) = \ln \lambda(u)$.  For the simulation study we fix $\gamma = 1/2$.

This model can be specified in \pkg{inlabru} as follows:
\begin{Schunk}
\begin{Sinput}
R> # Fix value for gamma parameter
R> gam <- 0.5
R> 
R> # Define an N(0,1) prior for u
R> u_prior <- list(prec = list(initial = 0, fixed = TRUE))
R> cmp <- ~ 0 + lambda(rep(1, n), model = "iid", hyper = u_prior,
+                      marginal = bru_mapper_marginal(qexp, rate = gam))
R> 
R> # Define model formula
R> fml <- y ~ log(lambda)
R> 
R> # Construct likelihood
R> lik <- like(
+    formula = fml,
+    family = "poisson",
+    data = ...
+  )
R> 
R> # Fit the model
R> fit <- bru(cmp, lik)
\end{Sinput}
\end{Schunk}
Note that the objects \code{gam} ($\gamma$ in Equation~\eqref{eq-poisson-exp}) and \code{n} (sample size) are stored in the global environment.  The component definitions and predictor formula can reference these objects directly.  Defining a parameter with a standard normal prior is done in the same way as with INLA, via the hyper option (see \code{?INLA::f}).
From \pkg{inlabru} version 2.10.0, the \code{bru\_mapper\_marginal} mapper class can be added to the component mapper pipeline directly, via the \code{marginal} argument.
The supplementary code includes comments about how to apply the transformation manually as well, using the internal helper function \code{bru\_forward\_transformation}, that transforms standard normal variables to given distributions using numerically stable methods.

\begin{figure}[!htb]
	\centering
	\includegraphics[width=\linewidth]{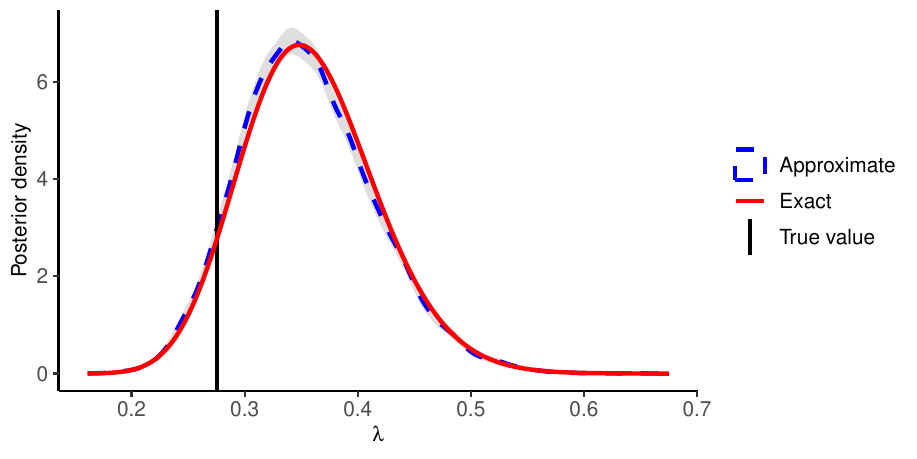}
	\caption{The approximate posterior density for $\lambda$ (dashed blue curve) compared to the true posterior density (solid red curve) for a single simulation.  The approximate posterior is a kernel smoothed representation of 20,000 posterior Monte Carlo samples for $\lambda$, with the shaded region showing 99\% pointwise bootstrap confidence intervals for the kernel density estimator.
	The true data-generating $\lambda$ values is marked with a solid black vertical line}
	\label{fig-toy-single}
\end{figure}

Figure~\ref{fig-toy-single} shows close agreement between the approximate posterior density estimated by inlabru alongside the exact posterior.  The approximate posterior density was calculated by Gaussian kernel smoothing 20,000 Monte Carlo samples from the posterior for $\lambda$.

To evaluate the ability of the approximate iterative INLA method to accurately estimate the posterior across many simulated datasets, we use an approach adapted from \cite{talts_ValidatingBayesianInference_2020}.  The basic idea is to repeatedly sample data from the prior predictive model, approximate the posterior for each sample and then calculate the posterior CDF value of each quantity of interest.  If the inference procedure accurately estimates the posterior then these CDF values will have a uniform distribution. See Supplementary Material Section~\ref*{sec:supplement:sbc} for more details.

For all simulations we set $\gamma = 0.5$ and sampled 100 values from the prior predictive distribution for $y$.  We ran 500 simulations and for each we computed the empirical CDF (ECDF) value for $\lambda(u)$ based on 5,000 Monte Carlo samples from the approximate posterior.

\begin{figure}[!htb]
  \centering
	\includegraphics{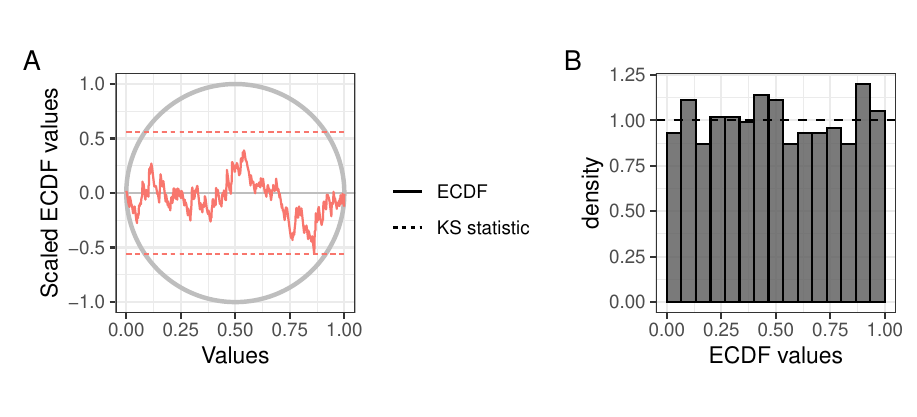}
	\caption{Results from the simulation based calibration checking for uniformity. Both plots are based on the ECDF values for $\lambda(u)$. \textbf{A}: Scaled ECDF values with the Kolmogorov-Smirnov test statistic; \textbf{B}: Histogram of ECDF values.}
	\label{fig-toy-sim}
\end{figure}

This shows that the approximate non-linear method is close to estimating the true posterior.  These figures show only a relatively small deviation from uniformity.  With such an approximate method based on a first-order Taylor expansion we should generally expect some deviation.  However, overall the approximation is fairly good for this model.


\section{Examples}
\label{examples}

We present three examples analysing a simulated dataset of counts in areal units.  Areal units are a common data structure in epidemiology and public health due to data commonly reported within administrative geographic units.  For the examples we use the intermediate zones \citep{IntermediateZoneBoundaries_2021} in the city of Glasgow, Scotland, which are freely available under the Open Government Licence \citep{OpenGovernmentLicence}.  These are areal units created for reporting data at medium sized geographic units, typically with 2,500-6,000 residents.

We simulated count data within each intermediate zone by assuming a dependence on a latent continuously indexed random field. We generated a single realisation from a Gaussian random field (GRF) with Mat\'ern covariance which we denote $\xi(\bm{s})$. This was then aggregated up to the areal unit level so that the count data in each areal unit $i$ were simulated from a Poisson distribution with rate $\lambda_i = \int_{\Omega_i} \exp(\beta + \xi(\bm{s})) \md\bm{s}$, where $\beta$ is an intercept parameter and $\Omega_i$ is the region of space covered by areal unit $i$.
In all three examples the data likelihood at the area level is Poisson with a log link function.  The three examples differ in the approach to modelling the log-rate for the Poisson variables.  We present 1) a Besag-York-Molli\'e (BYM)\citep{besag_BayesianImageRestoration_1991} example, 2) an example aggregating a continuously indexed random field to areal units, and 3) a joint likelihood model where the latent random field is observed at a set of point locations and aggregated to areal units.

\subsection{The Besag-York-Molli\'e model for areal data}

The Besag-York-Molli\'e model is an example of an intrinsic random field defined on a graph that represents the neighbourhood structure of the areal units.  The BYM component consists of the sum of a spatially structured intrinsic conditional autoregressive (ICAR) field $\bm{u}$ and a spatially unstructured Gaussian field $\bm{v}$. i.e. The log rate is $\ln \lambda_i = \beta + w_i$, where $w_i = u_i + v_i$.

The data object \code{glasgow} is an \pkg{sf} polygons object with the response variable named \code{count} and a variable named \code{ID} that is an integer identifier for each areal unit.
The model components and formula are
\begin{Schunk}
\begin{Sinput}
R> cmp <- ~ 0 + beta(1) + w(ID,
+    model = "bym",
+    graph = g
+  )
R> 
R> fml <- count ~ beta + w
\end{Sinput}
\end{Schunk}

where \code{g} is the graph defining the neighbourhood structure.
Note that we have chosen the component names \code{beta} and \code{w} to directly match the mathematical notation above.  To match the formula conventions of other packages, \pkg{inlabru} will include an intercept by default unless \code{+0} or \code{-1} is included in the components definition,
or there is a component name that begins with the string \code{intercept} or \code{Intercept}.

To fit and predict from the model:
\begin{Schunk}
\begin{Sinput}
R> lik <- like(
+    formula = fml,
+    data = glasgow,
+    family = "poisson"
+  )
R> 
R> fit <- bru(
+    components = cmp,
+    lik
+  )
R> 
R> predict(
+    fit,
+    newdata = glasgow,
+    formula = ~ exp(beta + w),
+    n.samples = 500
+  )
\end{Sinput}
\end{Schunk}

\begin{figure}[!htb]
	\centering
	\includegraphics{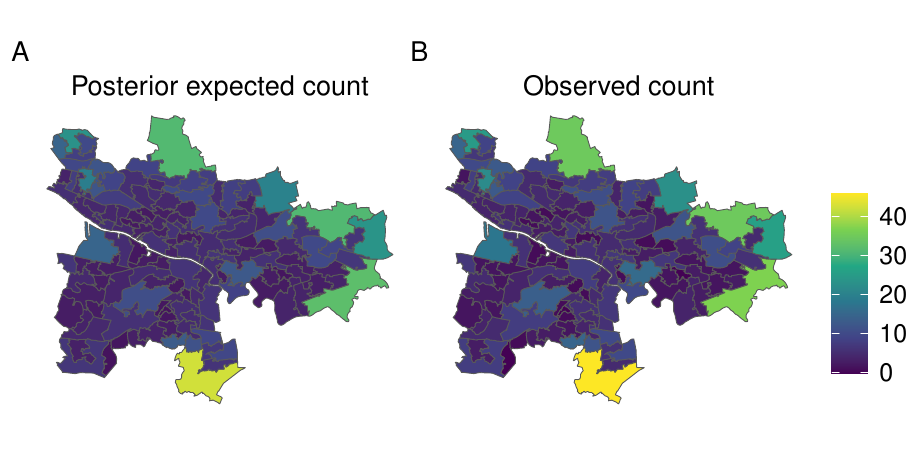}
	\caption{\textbf{A}: The BYM model posterior rate per area unit; \textbf{B}: The observed count per area unit.}
	\label{fig-bym}
\end{figure}

Figure~\ref{fig-bym} shows the mean posterior rate returned by the \code{predict()} call alongside the true rate used to simulate the data.

\subsection{Aggregating a continuously indexed random field to areal units}
\label{ex-agg}

This example demonstrates that the extra flexibility provided by the iterated INLA method allows for models to be fitted that are not possible to fit using standard INLA.

In this example we specify a continuously indexed latent random field using the SPDE approach.  We use the term continuously indexed to distinguish from the discrete area models like the BYM model.  In this case the index is a spatial location $\bs$ which can be any location in a continuous spatial domain.

This model structure is useful for applications where the response variable is available at a different spatial scale to the assumed causes.  For example, one might expect counts of asthma cases in an areal unit to depend on air pollution, which varies continuously in space and does not recognise areal unit boundaries.

The log-rate is approximated using a numerical integration scheme
$$
\begin{aligned}
\ln \lambda_i &= \ln \left( \int_{\Omega_i} \exp(\beta + \xi(\bm{s}))  \md\bm{s} \right) \\
               &\approx \ln \left[ \sum_j w_{ij} \exp(\beta + \xi(\bm{s}_{ij})) \right]
\end{aligned}
$$
with weights $w_{ij}$ defined using projected integration weights to mesh nodes.
In order to implement this, one could store these weights in a matrix $\bm{W}$
such that $\bm{\lambda} \approx \bm{W} \exp(\beta + \bm{\xi})$, using a
predictor expression written as \code{log(W \%*\% exp(Intercept + field))}. However, this procedure is already encapsulated in the \code{bru\_mapper\_logsumexp} mapper, so that the $\bm{W}$ matrix is constructed automatically and doesn't need to be visible in the user code. In addition, the
mapper uses an internal weight shift to avoid numerical over/underflow when summing the terms.
The code to construct the integration scheme can be found in Supplementary Material Section~\ref*{sec:supplement:aggregation}.

Given this integration scheme, the code to specify and fit the model is
\begin{Schunk}
\begin{Sinput}
R> # Define Matern/SPDE model
R> matern <- inla.spde2.pcmatern(
+    mesh,
+    prior.range = c(2, 0.1),
+    prior.sigma = c(3, 0.1)
+  )
R> 
R> # Define the intermediate zone integration scheme
R> ips <- fm_int(mesh, samplers = glasgow)
R> agg <- bru_mapper_logsumexp(
+    rescale = FALSE,
+    n_block = nrow(glasgow)
+  )
R> 
R> # Define model components and formula
R> cmp <- ~ 0 + beta(1) + xi(main = geometry, model = matern)
R> fml <- count ~ ibm_eval(
+    agg,
+    input = list(weights = weight, block = .block),
+    state = beta + xi
+  )
R> 
R> # Construct the likelihood
R> lik <- like(
+    formula = fml,
+    family = "poisson",
+    data = ips,
+    response_data = glasgow
+  )
R> 
R> # Fit the model
R> fit <- bru(
+    components = cmp,
+    lik
+  )
\end{Sinput}
\end{Schunk}
In this example we make use of the \code{response\_data} argument of the \code{like()} function, which allows users to specify response data of different length to what would usually be expected.  This feature of \pkg{inlabru} allows covariates and latent parameters to be on a different scale to the observed data.
The non-linear predictor function provides the link between these different scales. In this case the link is the numerical integration of the latent field within each intermediate zone where \code{ibm\_eval(agg, ...)} results in a vector of length $n$ that matches the response data.

The \code{xi} component is evaluated at the locations provided in
\code{ips}, which define the integration scheme.  Note that this is an
\code{sf} object in this example, and the \code{xi} component was defined using
\code{geometry} as input.  If \code{geometry} is not a named variable then
the software assumes it is a general R expression to be applied to the data.
In this case \code{geometry} is the geometry information column of the
\code{sf} data object, which is supported by the \code{bru\_mapper()}
method for the \code{matern} model object, that constructs the model matrix
for the component.  For old \code{SpatialPointsDataFrame} objects, one would
instead use \code{xi(coordinates, ...)}, where \code{coordinates} is the
\pkg{sp} function that extracts a 2-column matrix of coordinates from the
data object. However, this would drop all coordinate system projection
information, which is retained in the \code{sf} object, and so we recommend
using \code{sf} objects where possible to avoid associating data with
incorrect model locations.

The mapper \code{bru\_mapper\_logsumexp} here defines the link from the state vectors to the effects vector.  In other words, it is the logarithms of the weighted sums of the element-wise exponential of the state vectors \code{beta} and \code{xi}, taking into account of the input vectors \code{weight} and \code{.block}. \code{.block} argument gives the block index vector from the \code{fm\_int} function.  By default, the \code{rescale = FALSE} argument gives the plain weighted sums. With \code{rescale = TRUE}, the sum is rescaled by the sum of the weights within each block, as
$$
\ln\left[\frac{\sum_{j} w_{ij} \exp \big(\beta + \xi (\bm{s}_{ij}) \big)}{\sum_{j} w_{ij}}\right].
$$
The \code{n\_block} argument specifies a predetermined number of output blocks.
For more information, please see the theory and technical documentation on \code{bru\_mapper} system \citep{lindgren_mapper_2022}.

\subsubsection{Non-linear convergence assessment}

Note that because the predictor is non-linear in the parameters, this model is fitted using the iterated INLA method.  This non-linearity is assumed unless the user specifies the formula as \code{y ~ .} (see Section~\ref{sec-intro-ex}, for example), or provides no formula at all.

For assessing the convergence of the linearisation, we can call the function
\begin{Schunk}
\begin{Sinput}
R> bru_convergence_plot(fit)
\end{Sinput}
\end{Schunk}
\begin{figure}[!htb]
	\centering
	\includegraphics[width=1\linewidth]{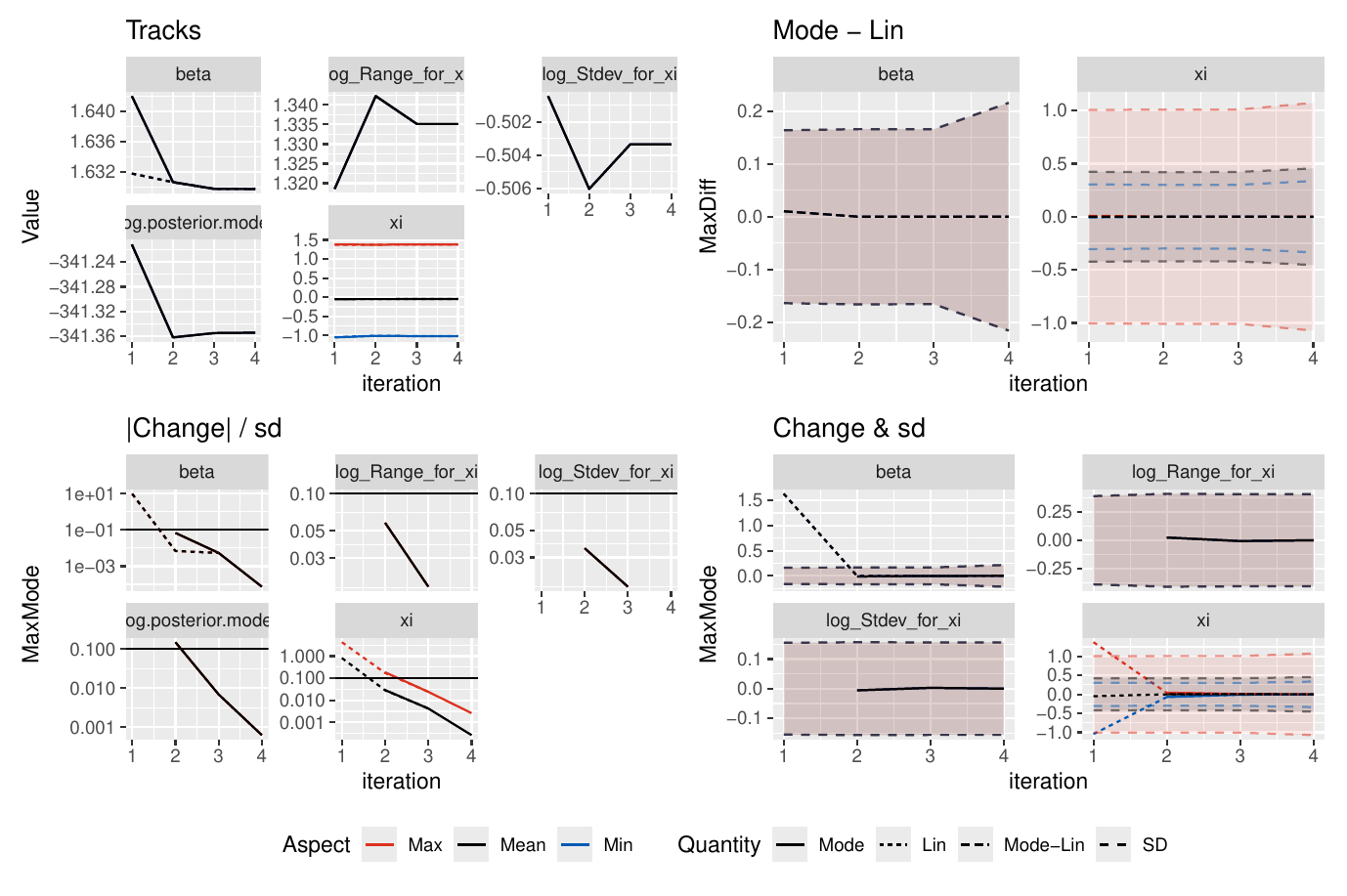}
	\caption{Figure for assessing the convergence of iterative INLA for Example \ref{ex-agg}.  This figure is explained in detail in the main text.}
	\label{fig-spatial-ex-2a}
\end{figure}
The resulting Figure~\ref{fig-spatial-ex-2a} shows the convergence of the linearisation across iterations, through four panels (from top to bottom, left to right),
\begin{itemize}
  \item \textit{Tracks}: The convergence of the modes of the covariates and hyperparameters.
  \item \textit{Mode - Lin}: The difference between the mode and the linearisation.
  \item \textit{$|$Change$|$/sd (Max and Mean)}: The absolute changes over the standard deviations (their maxima and means for vector components) for the mode and the linearisation.
  \item \textit{Change \& SD}: The changes and standard deviations of the covariates and hyperparameters.
\end{itemize}
In this case the convergence is virtually immediate, with the second iteration already being close enough to the first iteration to trigger then of iterations. The third and final step
only carries out the posterior integration step of the INLA method.

\subsubsection{Predictions}

Predictions can be generated at multiple scales by modifying the formula used in the \code{predict()} call.  For predictions at the areal unit level we again make use of the integration scheme in the predict formula:
\begin{Schunk}
\begin{Sinput}
R> # areal unit predictions using logsumexp mapper
R> fml_latent <- ~ ibm_eval(
+    agg,
+    input = list(weights = weight, block = .block),
+    state = beta + xi,
+    log = FALSE
+  )
R> predict(
+    fit,
+    newdata = ips,
+    formula = fml_latent
+  )
\end{Sinput}
\end{Schunk}

In addition to the component effects, the prediction method also provides access to the latent parameters directly, with the syntax \textit{<componentname>}\code{\_latent} appended to component names is used to access the latent parameters directly, with no need to pass a \code{newdata} argument to \code{predict()}. Similarly, the \textit{<componentname>}\code{\_eval(}\textit{<values>}\code{)} can be used to directly evaluate a component effect at given values of the inputs.

For predictions at a higher-resolution, we pass an \code{sf} object defining the locations at which we wish to make predictions.  For example,
\begin{Schunk}
\begin{Sinput}
R> # pixel level predictions
R> pts <- fm_pixels(mesh, mask = bnd)
R> pred <- predict(
+    fit,
+    newdata = pts,
+    formula = ~ exp(beta + xi)
+  )
R> pts$lambda_mean <- pred$mean
\end{Sinput}
\end{Schunk}
uses the \code{fm\_pixels()} function that generates an \code{sf} object that can be used as prediction locations.
The \code{mask} argument is optional and can be used to restrict the prediction locations to a specific region.  The \code{predict()} function will automatically detect the components involved in the predictor expressions and generate predictions at the appropriate locations.
This ability to use general \proglang{R} expressions in component definitions allows the software to be easily extended to new data structures.

\begin{figure}[!htb]
	\centering
	\includegraphics{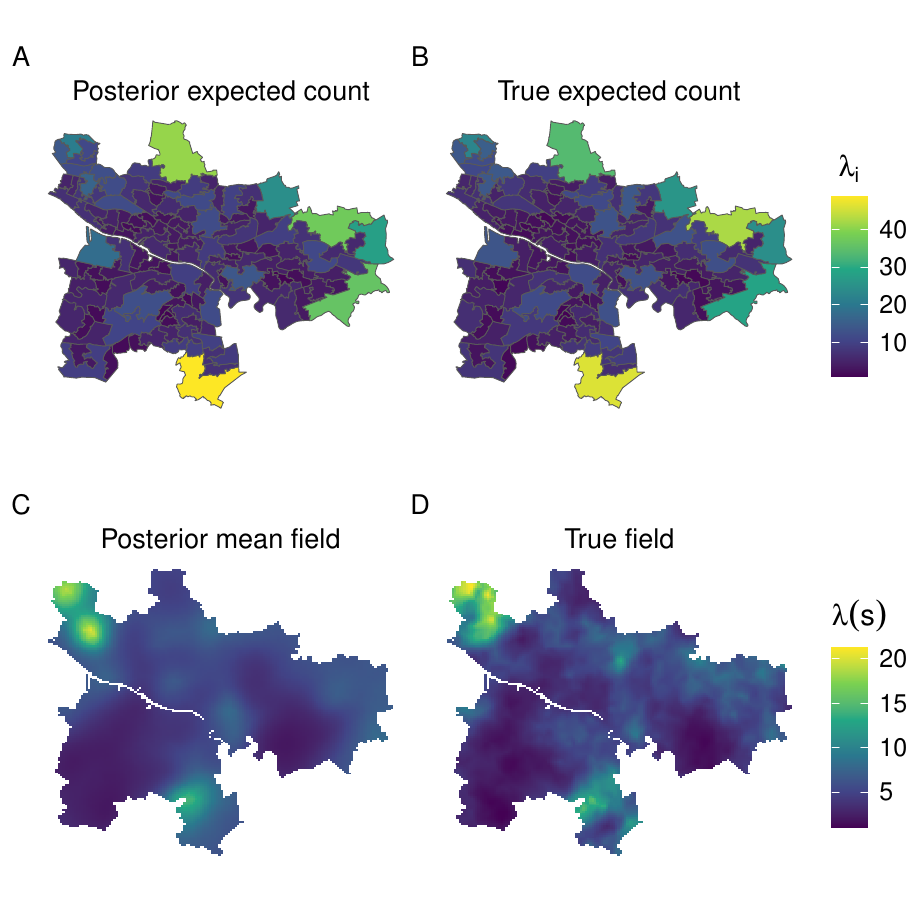}
	\caption{Summary of the fitted model from the aggregation Example \ref{ex-agg}. \textbf{A}: the posterior expected count at the area unit level; \textbf{B}: the true expected count at the area unit level; \textbf{C}: the posterior mean of the intensity field; \textbf{D}: the true intensity surface.}
	\label{fig-spatial-ex-2b}
\end{figure}

The posterior mean for both the areal unit rates and the pixel level predictions are shown in Figure~\ref{fig-spatial-ex-2b}, alongside the true values used to simulate the data.

\subsection{A joint model aggregating a continuously indexed random field to areal units with the field observed at point locations}
\label{ex-agg-joint}

This example uses the same aggregation scheme as in the above but also assumes that we have additional data observing the latent random field at a set of point locations.  This data structure can be common where information on relevant covariates is only available at a finite set of locations in the region of interest, such as measures of air pollution at air pollution monitoring stations, for example.
In addition to the count data at the areal unit level, we have data $z_k$ for $k = 1, \ldots, K$, and we assume the following model:
\begin{align*}
  z_k &= \alpha_0 +  \xi(\bm{s}_k) + \epsilon_k\\
  \ln \lambda(\bm{s}) &= \beta_0 + \beta_1 \xi(\bm{s}) \\
  \lambda_i &\approx \sum_j \ln \left(w_{ij} \lambda(\bm{s}_{ij}) \right) , \\
\end{align*}
where $\alpha_0$ is an intercept parameter which could account for systematic measurement or calibration error involved in measuring the random field $\xi(\bm{s})$ and the $w_{ij}$ are defined as above.  We do not include $\alpha_0$ to the second line to avoid identifiability issues. The likelihood for $\bm{z}$ is Gaussian which also incorporates iid measurement error.  In this example we view the random field as though it were a fixed effect covariate that we have only observed in a finite set of locations.  This partially observed covariate, which we model using the SPDE effect, affects the continuously indexed log-rate linearly through the parameter $\beta_1$.

\begin{figure}[!htb]
	\centering
	\includegraphics{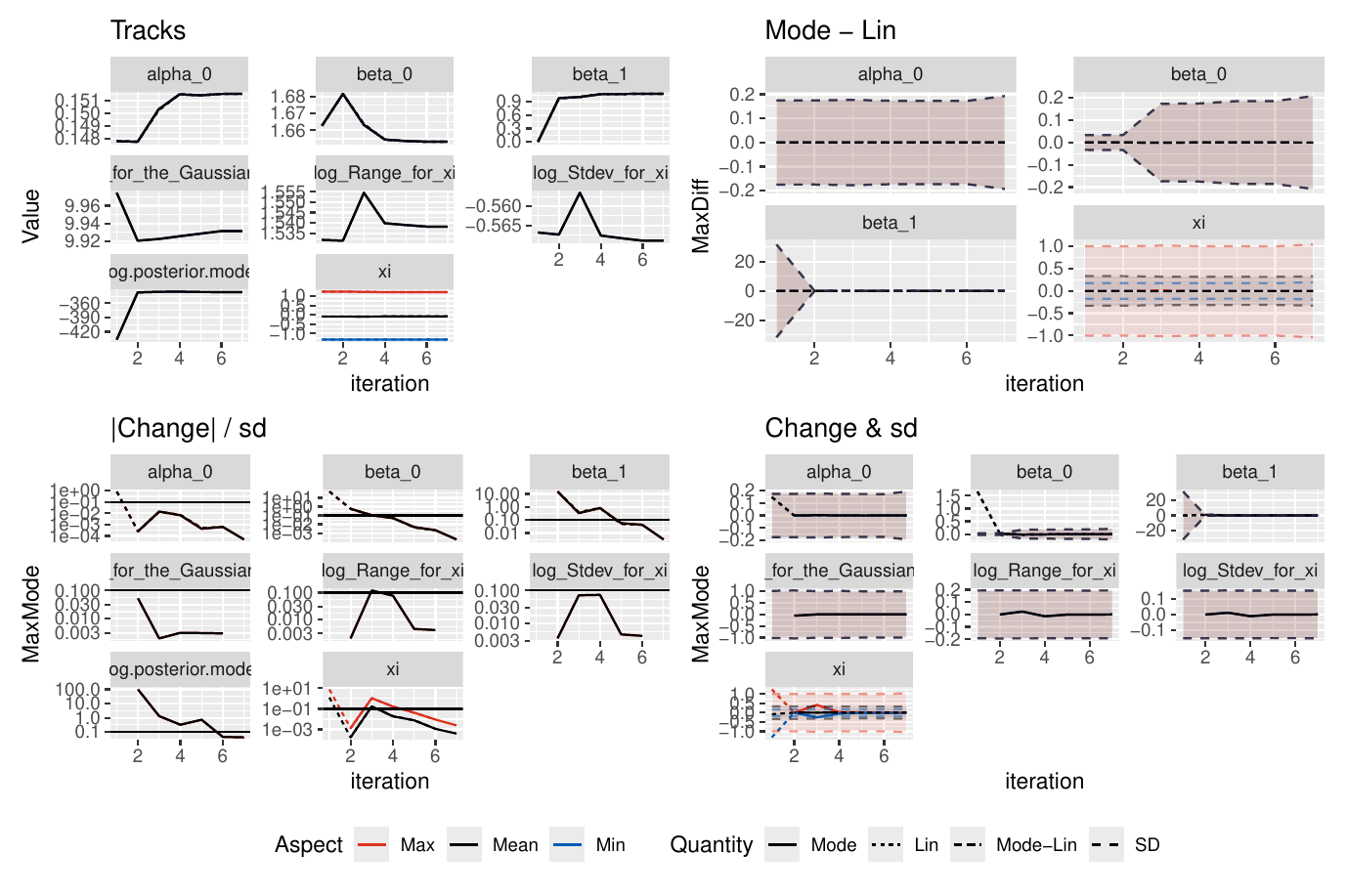}
	\caption{Convergence diagnostic plot for the joint model Example \ref{ex-agg-joint}.}
		\label{fig-spatial-ex-3-convergence}
\end{figure}

The code to specify and fit the model is
\begin{Schunk}
\begin{Sinput}
R> # Define model components
R> cmp <- ~ 0 + xi(main = geometry, model = matern) +
+    alpha_0(1) +
+    beta_0(1) +
+    beta_1(1)
R> 
R> # Gaussian likelihood for z observations
R> z_fml <- z ~ alpha_0 + xi
R> z_lik <- like(
+    "Gaussian",
+    formula = z_fml,
+    data = z_data
+  )
R> 
R> # Poisson likelihood for count observations
R> count_fml <- count ~ ibm_eval(
+    agg,
+    input = list(weights = weight, block = .block),
+    state = beta_0 + beta_1 * xi
+  )
R> count_lik <- like(
+    "Poisson",
+    formula = count_fml,
+    data = ips,
+    response_data = glasgow
+  )
R> 
R> # Fit joint likelihood model
R> fit <- bru(
+    components = cmp,
+    z_lik,
+    count_lik,
+    options = list(
+      bru_max_iter = 10,
+      bru_verbose = 2,
+  #    bru_initial = list(
+  #      beta_1 = 1
+  #    )
+    )
+  )
\end{Sinput}
\end{Schunk}
where \code{z\_data} is an \code{sf} object with a column named \code{z}.
Predictions can be generated from the model in a similar way to the previous examples.
The methods will normally auto-detect which effects are involved in the predictor expressions,
but if needed, the default auto-detection can be overridden using the \code{used}
argument.  We can optionally include the \code{bru\_initial} option to set the initial values for the latent variables, to potentially speed up the method convergence, see Figure~\ref{fig-spatial-ex-3-convergence}. We believe that the random field $\xi(\bm{s})$ would contribute positive effect on the intensity $\lambda (\bm{s})$.  The \code{bru\_verbose} option controls the amount of information about the iterative algorithm that is printed by \code{bru()}.  The verbose information for this model is included in Supplementary Material Section~\ref*{sec:supplement:verbose}.

A seen by the convergence diagnostic plot in Figure~\ref{fig-spatial-ex-3-convergence},
the iterated linearisation method converges in 6 steps, with the final 7th
step only computing the posterior integration step of the INLA method.

\begin{figure}[!htb]
	\centering
	\includegraphics{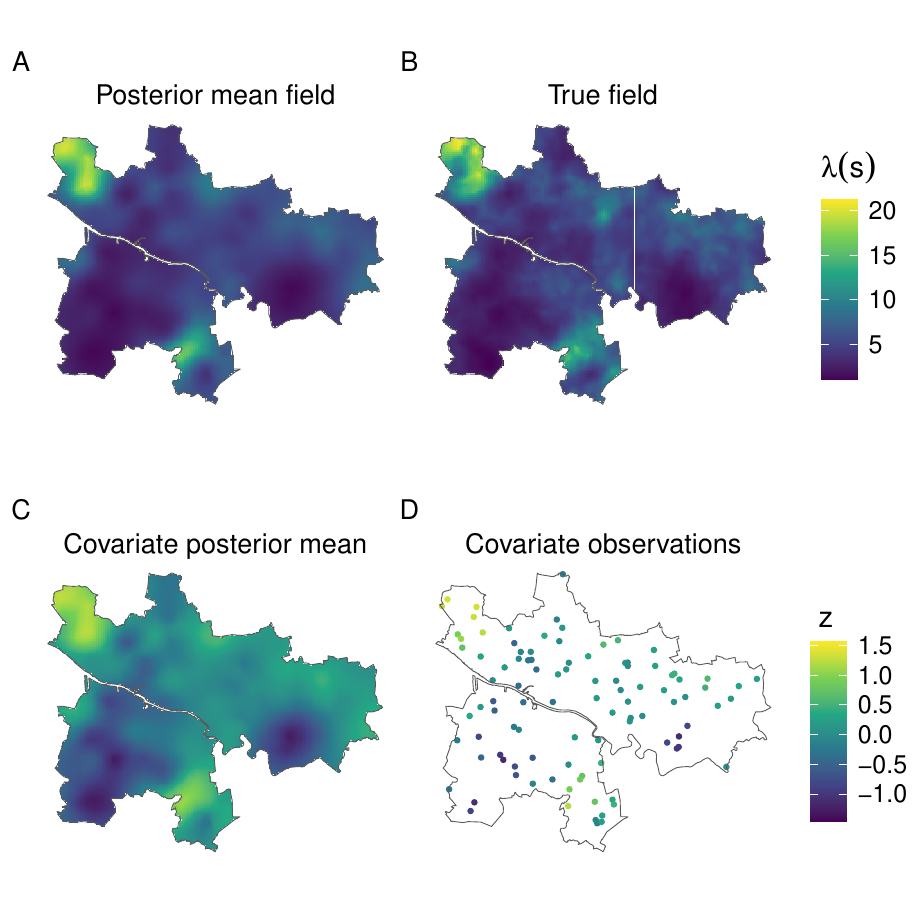}
	\caption{Summary of the fitted model for the aggregation joint likelihood model in Example \ref{ex-agg-joint}.  \textbf{A}: the posterior mean of the intensity field; \textbf{B}: the true intensity surface; \textbf{C}: the posterior mean of the covariate field; \textbf{D}: the observations of the covariate field.}
		\label{fig-spatial-ex-3}
\end{figure}

The posterior mean for $\lambda(\bm{s})$ and for $\xi(\bm{s})$ are shown in Figure~\ref{fig-spatial-ex-3}, alongside the truth and the latent field observations.

\section{Discussion}
\label{discussion}


\subsection{Other methods for extending INLA}

The iterative INLA approach extends the class of models that can be fitted using INLA.  Other research has investigated this by embedding INLA within other inference frameworks such as Metropolis Hastings Markov chain Monte Carlo (MH-MCMC) \citep{gomez-rubio_MarkovChainMonte_2018} and importance sampling \citep{berild_ImportanceSamplingIntegrated_2022}.  These approaches allow users to fit conditional latent Gaussian models, where the model is, conditional on certain parameters, able to be fitted using standard \pkg{INLA}.

In the case of MH-MCMC, this comes at considerable extra computational cost.  Typical chains can require fitting a model using \pkg{INLA} tens or even hundreds of thousands of times, which is prohibitively expensive for all but the most simple of models.  This computational cost has hindered embedding INLA within MCMC approaches.  In our experience so far, iterative INLA generally requires something between a couple and twenty iterations to converge, with simpler models often converging in 3 or 4 steps.

The class of models that can be fitted using \pkg{inlabru} is smaller than the class of conditional LGMs because only the latent predictor expression is allowed to be non-linear.  However, the data likelihood must still be one of those that is currently implemented in \pkg{INLA}.  In contrast, the MH-MCMC approach is, in theory at least, not restricted in this sense.

An alternative to MH-MCMC is to use importance sampling (IS) to marginalise over the conditioning parameters for the conditional LGM.  Let $\theta_{c}$ denote the parameters conditional on which the other model parameters, $\theta_{-c}$, can be estimated using INLA.  Then the unconditional marginal posterior distributions for elements of $\theta_c$ can be obtained by
$$
\pi(\theta_{-c,j}) | \bm{y}) = \int \pi(\theta_{-c,j} | \bm{y}, \theta_c) \pi(\theta_c | \bm{y}) \md\theta_c.
$$
\cite{berild_ImportanceSamplingIntegrated_2022} present estimating this integral using IS.  The importance sampling weight can also be used to estimate the joint posterior $\pi(\theta_c | \bm{y})$.  The performance of IS is sensitive to the choice of proposal distribution.  The greater the difference from $\pi(\theta_c | \bm{y})$, the more samples are required.  To reduce the risk of a poor choice, \cite{berild_ImportanceSamplingIntegrated_2022} suggest using an adaptive IS algorithm.  This approach appears more promising than MH-MCMC, with examples showing good convergence to true posterior in under 30 adaptations of the proposal distribution.  However, the examples in their paper are more simple than those presented here.

\cite{stringer_FastScalableApproximations_2023} present a method inspired by INLA for efficient fitting of what they term extended LGMs.  They propose an alternative approach to inference that, unlike the original INLA implementation, avoids including the predictor as part of the latent field and use this feature to do aggregation similar the spatial examples in this paper.  Key to their approach is flexibility in specifying the model likelihood.  Whereas \pkg{inlabru} relies on the likelihoods implemented in the core \pkg{INLA} C code, their software can work with a general \proglang{C++} template of the likelihood, which gives users flexibility to describe non-linear relationships.  Since the new variational Bayes method in \pkg{INLA} similarly doesn't include the the predictor in the latent field, this user-plugin feature is the main practical benefit of the alternative approach.

\subsection{Examples of inlabru in action}

Several papers have been published that use the non-linear predictor feature of \pkg{inlabru}. \cite{arceguillen_AccountingUnobservedSpatial_2023} use \pkg{inlabru} to analyse telemetry data from GPS tags on wild animals.  They use the non-linear predictor feature to model animal movement kernels and formulate the likelihood as a Cox process (which is not implemented in INLA, but is in \pkg{inlabru}, see \code{?lgcp}) to include spatially correlated random effects for the first time in this class of model.

\cite{serafini_ApproximationBayesianHawkes_2023} use \pkg{inlabru} to fit the self-exciting Hawkes point process.  In these models the excitation function is non-linear and so cannot be fitted using INLA.  This work has been implemented in the \pkg{ETAS.inlabru} package \citep{naylor_BayesianModelingTemporal_2023} for modelling seismic activity using point process models.

\cite{martino_IntegrationPresenceonlyData_2021} use \pkg{inlabru} to combine various opportunistic data sources to model the abundance and spatial distribution of dolphins.  They use the non-linear predictor to model the decreasing likelihood of observing dolphins further away from tourist hotspots or vessels through a half-normal detection function.

All of the above examples are models with spatial or temporal dependence where researchers would like to be able to capitalise on the computational efficiency offered by the INLA method.  However, due to the nature of the analyses, these models cannot be fitted using standard \pkg{INLA} package implementation.
Using the iterative INLA approach extends the capability of researchers to address their research questions.

\subsection{Software that uses inlabru}

In addition to \pkg{ETAS.inlabru} already mentioned above there are several pieces of software that depend on \pkg{inlabru}. The \pkg{dirinla} package \citep{dirinla2022} implements Dirichlet regression for analysing compositional data using INLA.  The \pkg{rSPDE} package \citep{rSPDE2023, bolin_CovarianceBasedRational_2024} is software for computing rational approximations of fractional SPDEs, which allows for inference on the fractional parameter in INLA SPDE models \citep{lindgren_ExplicitLinkGaussian_2011}.  The \pkg{MetricGraph} package \citep{bolin_metric_2024, MetricGraph2023} is software for fitting models with random fields defined on graphs. \pkg{INLAspacetime} \citep{Krainski_INLAspacetime_2023} supports spatio-temporal modelling using the \code{cgeneric} interface for defining new model components and supports non-separable space-time models  \citep{lindgren_DiffusionbasedSpatiotemporalExtension_2023}.  The \pkg{fdmr} package \citep{fdmr2023} supports interactive spatio-temporal modelling and includes features for defining spatially explicit model components, choosing priors, and evaluating model outputs.

\subsection{Future improvements}

There are several avenues for improvements to \pkg{inlabru} moving forwards.  The implementation of the KL-divergence result (Equations \eqref{eq-KL-1} and \eqref{eq-KL-2}) is feasible but requires further development time.  The \code{like()} function has the potential to be improved by adding modularity and extracting certain model building steps into separate functions. This should allow for easier extensions to complex user-defined observation models, such as specific aggregation models, as well as support for add-on packages.  Finally, although the interoperability with \pkg{INLA} is considerable, when new features are introduced to \pkg{INLA}, they in a small number of cases require special support to be implemented in \pkg{inlabru}.
An example of this are observation models where special data formats are required, such as \code{inla.mdata} and
\code{inla.surv}. From \pkg{INLA} version 24.06.27 and \pkg{inlabru} development version 2.10.1.9011,
\pkg{inlabru} can be used with both \code{inla.mdata} and \code{inla.surv} objects, so that they will be fully supported
when version 2.11.0 is released.

The results in the paper were generated with \pkg{INLA} 24.06.27, \pkg{inlabru} 2.10.1.9012, and \pkg{fmesher} 0.1.6.9003.

\bibliography{paper}

\appendix

\section*{Supplementary Information}

Since \pkg{inlabru} is under active development, some information in the paper and supplementary materials may become outdated.  We encourage readers to check the articles on the \pkg{inlabru} website for up to date information.  The articles can be accessed here:

\url{https://inlabru-org.github.io/inlabru/articles/}

\section{Mapper information}
\label{sec:supplement:mappers}

The \code{bru\_mapper} methods define model component design matrices are constructed based on the input information.  This is a summary of the \code{bru\_mapper} methods, a more detailed version can be found in the \code{bru\_mapper} vignette on the \pkg{inlabru} website \citep{lindgren_mapper_2022} as well as in documentation \code{?bru\_mapper} and \code{?bru\_mapper\_methods}.

When implementing new user-defined model components, or models defined in add-on packages, e.g.\ via the \code{rgeneric} nd \code{cgeneric} \pkg{INLA} frameworks, a new mapper class can be defined, with a \code{bru\_get\_mapper} method that \pkg{inlabru} can call to automatically obtain a suitable mapper object.

\subsection{Mapper system introduction}
\label{sec:supplement:mapper-intro}

Each inlabru latent model component generates an \textit{effect}, given the latent \textit{state} vector,
i.e.\ a vector of values for the component.
The purpose of the mapper system is to define the link from state vectors to effect vectors.
For an ordinary model component, named $c$, this link can be represented as a function of
the latent variables,
$$
\wt{\bm{\eta}}_c = f_c(\bm{u}_c, \text{input}_c),
$$
where $\bm{u}_c$ is the latent state vector, $\wt{\bm{\eta}}_c$ is the resulting \emph{component effect} vector, and
$\text{input}_c$ are fixed component parameters, and $f_c(\cdot)$ is a linear or non-linear transformation function, defined by a combination of mapper methods. Each component has
a linearised version, at a given linearisation state $\bm{u}_c^0$, defined by
the affine transformation
$$
\ol{\bm{\eta}}_c = \bm{b}^0_c(\text{input}_c) + \bm{A}^0_c(\text{input}_c) \bm{u}_c,
$$
where $\bm{b}_c(\text{input}^0_c)$ is a \emph{component offset}, and $\bm{A}^0_c(\text{input}_c)$ is a \emph{component design matrix}. This matrix depends on the component
inputs (covariates, index information, etc) from \code{main}, \code{group}, \code{replicate}, \code{weights}, and \code{scale} in the component definition, as well as the linearisation point $\bm{u}_c^0$. In addition a marginal transformation mapper can be applied, to convert
$\pN(0,1)$ marginal distributions into other, pre-specified distributions.

The in-package documentation for the mapper methods is contained in four parts:

\begin{Schunk}
\begin{Sinput}
R> ?bru_mapper # Mapper constructors
R> ?bru_mapper_generics # Generic and default methods
R> ?bru_mapper_methods # Specialised mapper methods
R> ?bru_get_mapper # Mapper extraction methods
\end{Sinput}
\end{Schunk}

Regular users normally at most need some of the methods from \code{?bru\_mapper}
and sometimes \code{?bru\_mapper\_generics}.
The methods in \code{?bru\_mapper\_methods} provides more details and allows the user
to query mappers and to use them outside the context of inlabru model definitions.
The \code{bru\_mapper\_define()} and \code{bru\_get\_mapper()} methods are needed for those
implementing their own mapper class.

\subsection{Mappers}

The main purpose of each mapper class is to allow evaluating a component effect, from given \code{input} and latent \code{state} ($\bm{u}_c$), by calling

\begin{Schunk}
\begin{Sinput}
R> ibm_eval(mapper, input, state)
\end{Sinput}
\end{Schunk}

for the \code{mapper} associated with the component definition.

\subsection{Basic mappers}

Basic mappers take covariate vectors or matrices as input and numeric vectors as state.

The basic constructors are:

\begin{itemize}
\item \code{bru\_mapper\_const()}
\item \code{bru\_mapper\_linear()}
\item \code{bru\_mapper\_index(n)}
\item \code{bru\_mapper\_factor(values, factor\_mapping)}
\item \code{bru\_mapper\_matrix()}
\item \code{bru\_mapper\_harmonics(order, scaling, intercept, interval)}
\end{itemize}

See the above-mentioned \code{bru\_mapper} vignette for the full definition of each mapper.

\subsection{Transformation mappers}

Transformation mappers are mappers that would normally be combined with other
mappers, as steps in a sequence of transformations, or as individual
transformation mappers.

\begin{itemize}
\item \code{bru\_mapper\_scale()}
\item \code{bru\_mapper\_marginal(qfun, pfun, ...)}
\item \code{bru\_mapper\_aggregate(rescale)}
\item \code{bru\_mapper\_logsumexp(rescale)}
\end{itemize}

Details on each is in the package documentation and the mapper vignette.

\subsection{Compound mappers}

Compound mappers define collections or chains of mappings,
and can take various forms of input. The state vector is normally a numeric vector,
but can in some cases be a list of vectors.

\begin{itemize}
\item \code{bru\_mapper\_collect(mappers, hidden)}
\item \code{bru\_mapper\_multi(mappers)}
\item \code{bru\_mapper\_pipe(mappers)}
\end{itemize}

Full details are in the package documentation and mapper vignette.

\subsection{Mapper methods}

Mapper objects themselves have methods associated with them to retrieve relevant information. Each mapper has the information required by \pkg{inlabru} to construct the full model matrix for the call to \pkg{INLA}.  The mapper methods are

\begin{Schunk}
\begin{Sinput}
R> ibm_n(mapper, inla_f, ...)
R> ibm_n_output(mapper, input, ...)
R> ibm_values(mapper, inla_f, ...)
R> ibm_jacobian(mapper, input, state, ...)
R> ibm_eval(mapper, input, state, ...)
R> ibm_names(mapper, ...)
R> ibm_inla_subset(mapper, ...)
\end{Sinput}
\end{Schunk}

For more information see the vignette.

\subsection{Defining new mappers}

Taken together, the basic constructors, transformation mappers and compound mappers offer functionality for users to define new mappers for their own model components.

A mapper object should store enough information in order for the \code{ibm\_\*} methods
to work. The simplest case of  a customised mapper is to just attached a new class
label to the front of the S3 \code{class()} information of an existing mapper,
to obtain a class to override some of the standard \code{ibm\_\*} method
implementations.

More commonly, the \code{bru\_mapper\_define()}
method should be used to properly set the class information:
\begin{Schunk}
\begin{Sinput}
R> bru_mapper_define(mapper, new_class)
\end{Sinput}
\end{Schunk}

For users wishing to write their own mappers we strongly suggest reading the mapper vignette which includes more information and provides examples of defining new mappers.

\section{Iterative method details}
\label{sec:supplement:iteration}

\subsection{Fixed point iteration}

The choice of the linearisation
point is key for the accuracy of the approximation.  To define a search and convergence criterion for choosing the linearisation point, we define a Bayesian estimation functional $f(\ol{p}_{\bm{v}})$ of the posterior
distribution linearised at $\bm{v}$.  This functional takes as input the posterior from a linearised model $\ol{p}_{\bm{v}}$ and outputs a new linearisation point.

We define the functional
$$
f(\ol{p}_{\bm{v}}) = (\wh{\bm{\theta}}_{\bm{v}},\wh{\bm{u}}_{\bm{v}})
$$
where
$$
\wh{\bm{\theta}}_{\bm{v}} = \argmax_{\bm{\theta}} \ol{p}_{\bm{v}} ( \bm{\theta} | \bm{y}),
$$
the posterior mode for $\bm{\theta}$, and
$$
\wh{\bm{u}}_{\bm{v}} = \argmax_{\bm{u}} \ol{p}_{\bm{v}} (\bm{u} | \bm{y}, \wh{\bm{\theta}}_{\bm{v}}),
$$
the joint conditional posterior mode for $\bm{u}$.  In other words, the functional generates a new linearisation point defined as the posterior mode for $\bm{\theta}$ and the conditional mode for $\bm{u}$ given $\bm{\theta}$.

An obvious choice for the optimal linearisation point is the posterior mode under the non-linear model.  Let $(\wh{\bm{\theta}},\wh{\bm{u}})$ denote the corresponding posterior modes for the true posterior distribution,
$$
\begin{aligned}
    \wh{{\bm{\theta}}} &= \argmax_{\bm{\theta}} p ( {\bm{\theta}} | \bm{y} ), \\
    \wh{\bm{u}} &= \argmax_{\bm{u}} p (\bm{u} | \bm{y}, \wh{{\bm{\theta}}}).
\end{aligned}
$$

We therefore seek a fixed point of the functional $(\bm{\theta}_*,\bm{u}_*)=f(\ol{p}_{\bm{u}_*})$ that ideally would be close to  $(\wh{\bm{\theta}},\wh{\bm{u}})$. We can achieve this for the conditional
latent mode, so that $\bm{u}_*=\argmax_{\bm{u}} p (\bm{u} | \bm{y}, \wh{\bm{\theta}}_{\bm{u}_*})$.  We therefore seek the latent vector $\bm{u}_*$ that generates the fixed point of the
functional, so that $(\bm{\theta}_*,\bm{u}_*)=f(\ol{p}_{\bm{u}_*})$.

One key to the fixed point iteration is that the observation model is linked to
$\bm{u}$ only through the non-linear predictor $\wt{\bm{\eta}}(\bm{u})$, since
this leads to a simplified line search method below.

\begin{enumerate}
\setcounter{enumi}{-1}
 \item Let $\bm{u}_0$ be an initial linearisation point for the latent variables
   obtained from the initial INLA call. Iterate the following steps for $k=0,1,2,...$
 \item Compute the predictor linearisation at $\bm{u}_0$.
 \item Compute the linearised INLA posterior $\ol{p}_{\bm{u}_0}(\bm{\theta}|\bm{y})$.
\item Let $(\bm{\theta}_1,\bm{u}_1)=(\wh{\bm{\theta}}_{\bm{u}_0},\wh{\bm{u}}_{\bm{u}_0})=f(\ol{p}_{\bm{u}_0})$ be the initial candidate for new
   linearisation point.
\item Let $\bm{v}_\alpha=(1-\alpha)\bm{u}_1+\alpha\bm{u}_0$, and find the value
   $\alpha$ minimises $\|\wt{\bm{\eta}}(\bm{v}_\alpha)-\ol{\bm{\eta}}(\bm{u}_1)\|$.
\item Set the new linearisation point $\bm{u}_0$ equal to $\bm{v}_\alpha$ and repeat from step 1,
   unless the iteration has converged to a given tolerance.
\end{enumerate}

A potential improvement of step 4 might be to also take into account the prior distribution for $\bm{u}$ as a minimisation penalty, to avoid moving further than would be indicated by a full likelihood optimisation.

\subsection{Line search details}
\label{sec:supplement:linesearch}

In step 4, we would ideally want $\alpha$ to be
$$
\argmax_{\alpha} \left[\ln p(\bm{u}|\bm{y},\bm{\theta}_1)\right]_{\bm{u}=\bm{v}_\alpha}.
$$
However, since this requires access to the internal likelihood and prior density
evaluation code, we instead use a simpler alternative. We consider norms of the form $\|\wt{\eta}(\bm{v}_\alpha)-\ol{\eta}(\bm{u}_1)\|$
that only depend on the nonlinear and linearised predictor expressions, and other known quantities, given $\bm{u}_0$, such as the current INLA estimate of the component wise predictor variances.

Let $\sigma_i^2 = \mathrm{Var}_{\bm{u}\sim \ol{p}(\bm{u}|\bm{y},\bm{\theta}_1)}(\ol{\bm{\eta}}_i(\bm{u}))$
be the current estimate of the posterior variance for each predictor element $i$.
We then define an inner product on the space of predictor vectors as
$$
\langle \bm{a},\bm{b} \rangle_V
=
\sum_i \frac{a_i b_i}{\sigma_i^2} .
$$
The squared norm for the difference between the predictor vectors $\wt{\bm{\eta}}(\bm{v}_\alpha)$
and $\ol{\bm{\eta}}(\bm{u}_1)$,with respect to this inner product, is defined as
$$
\| \wt{\bm{\eta}}(\bm{v}_\alpha) - \ol{\bm{\eta}}(\bm{u}_1)\|^2_V
=
\sum_i \frac{|\wt{\bm{\eta}}_i(\bm{v}_\alpha)-\ol{\bm{\eta}}_i(\bm{u}_1)|^2}{\sigma_i^2} .
$$
Using this norm as the target loss function for the line search avoids many potentially
expensive evaluations of the true posterior conditional log-density. We evaluate
$\wt{\bm{\eta}}_1=\wt{\bm{\eta}}(\bm{u}_1)$ and make use of the linearised predictor
information. Let $\wt{\bm{\eta}}_\alpha=\wt{\bm{\eta}}(\bm{v}_\alpha)$ and $\ol{\bm{\eta}}_\alpha=\ol{\bm{\eta}}(\bm{v}_\alpha)=(1-\alpha)\wt{\bm{\eta}}(\bm{u}_0)+\alpha\ol{\bm{\eta}}(\bm{u}_1)$.
In other words, $\alpha=0$ corresponds to the previous linear predictor, and $\alpha=1$
is the current estimate from INLA.
An exact line search would minimise $\|\wt{\bm{\eta}}_\alpha-\ol{\bm{\eta}}_1\|$.
Instead, we define a quadratic approximation to the
non-linear predictor as a function of $\alpha$,
$$
\breve{\bm{\eta}}_\alpha =
\ol{\bm{\eta}}_\alpha + \alpha^2 (\wt{\bm{\eta}}_1 - \ol{\bm{\eta}}_1)
$$
and minimise the quartic polynomial in $\alpha$,
$$
\begin{aligned}
\|\breve{\bm{\eta}}_\alpha-\ol{\bm{\eta}}_1\|^2
&=
\| (\alpha-1)(\ol{\bm{\eta}}_1 - \ol{\bm{\eta}}_0) + \alpha^2 (\wt{\bm{\eta}}_1 - \ol{\bm{\eta}}_1) \|^2
.
\end{aligned}
$$

If initial expansion and contraction steps are carried out, leading to an initial
guess of $\alpha=\gamma^k$, where $\gamma>1$ is a scaling factor (see \code{?bru\_options}, \code{bru\_method\$factor}) and $k$ is the
(signed) number of expansions and contractions, the quadratic expression is replaced by
$$
\begin{aligned}
\|\breve{\bm{\eta}}_\alpha-\ol{\bm{\eta}}_1\|^2
&=
\| (\alpha-1)(\ol{\bm{\eta}}_1 - \ol{\bm{\eta}}_0) + \frac{\alpha^2}{\gamma^{2k}} (\wt{\bm{\eta}}_{\gamma^k} - \ol{\bm{\eta}}_{\gamma^k}) \|^2
,
\end{aligned}
$$
which is minimised on the interval $\alpha\in[\gamma^{k-1},\gamma^{k+1}]$.

\section{Approximate Bayesian method checking}
\label{sec:supplement:sbc}

We take an approach adapted from \cite{talts_ValidatingBayesianInference_2020}.  Given a Bayesian hierarchical model of the form
$$
\begin{aligned}
	\theta &\sim p(\theta) \\
	u &\sim p(u | \theta) \\
	y &\sim p(y | u, \theta),
\end{aligned}
$$
the aim is to assess the ability of an approximate Bayesian inference method to estimate the posterior of some functional $h(\theta, u)$. An approximate posterior $\wt{p}(u, \theta | y)$ can be checked by simulating data from the exact model and checking approximate posterior CDF values for uniformity.

Given samples
$$
\begin{aligned}
	\theta^{(k)} &\sim p(\theta) \\
	u^{(k)} &\sim p(u | \theta^{(k)}) \\
	y^{(k)} &\sim p(y | u, \theta^{(k)}),
\end{aligned}
$$
for $k = 1, \ldots, K$, the CDF value is defined as
$$
	w^{(k)} = F(h(\theta, u)),
$$
where $F(.)$ is the CDF for $h(\theta, u)$ under the approximate posterior $\wt{p}(h(\theta, u) | y)$.  If the approximate inference method recovers the correct posterior distributions then $w^{(k)} \sim \pUnif(0,1)$, independently for $k = 1, \ldots K$.

If the $F(.)$ is not available in closed form then it can be estimated by samples from $\wt{p}(u, \theta | y)$.
For each $k = 1, \ldots, K$, sample $J$ times from $\wt{p}(u, \theta | y^{(k)})$ and calculate the empirical CDF value
$$
  \wt{w}^{(k)} = \frac{1}{J}\sum_{j=1}^J \mathbb{I}(h(\theta^{(j|k)}, u^{(j|k)}) < h(\theta^{(k)}, u^{(k)})) - \frac{1}{2J}.
$$
The empirical CDF value $\wt{w}^{(k)}$ is a normalised rank statistic that, if the approximate inference approach recovers the posterior exactly, will have a discrete uniform distribution on an equal partition of $(0, 1)$.  Note that this is an adjusted form to that given in \cite{talts_ValidatingBayesianInference_2020}, which works directly with the rank statistic $\sum_{j=1}^J \mathbb{I}(h(\theta^{(j|k)}, u^{(j|k)}) < h(\theta^{(k)}, u^{(k)}))$ which is uniform on the integers $\{0, 1, \ldots, J\}$.  The adjusted form transforms this to the interval $(0, 1)$, not inclusive of the endpoints.  Choosing $J$ to be large allows samples to be approximately evaluated using standard approaches for checking samples from a continuous uniform distribution, such as the Kolmogorov-Smirnov test, which we use in Section~\ref{method-checking}.

As noted by \citet{modrak_simulation-based_2023}, the calibration checking approach above is insensitive to certain types of deviations from the true posterior distribution. The solution suggested there, including assessing functionals that involve the observation log-likelihoods, may also be implemented for \pkg{INLA} and \pkg{inlabru} results.

\section{Setting priors}

It is important to consider priors carefully for non-linear model components.  For example, consider the default INLA prior of a \code{"iid"} Gaussian model $x$, which is mean zero and precision equal to $5 \times 10^{-5}$, which is equivalent to a standard deviation of roughly 141.  Then $\exp(x)$ will have a variance that is larger than the typical maximum number that can be represented in the double-precision floating point number format that \proglang{R} uses.  Care should therefore be taken when setting priors for non-linear predictors. As is the case in \pkg{INLA}, priors can be set by using the \code{hyper} argument when defining model components.

\section{Aggregation example integration scheme}
\label{sec:supplement:aggregation}

In the examples in Section~\ref{ex-agg} of the paper, the predictor consists of a numerical integration scheme applied to the latent GMRF parameters:
$$
\begin{aligned}
\ln \lambda_i &= \ln \left( \int_{\Omega_i} \exp(\beta + \xi(\bm{s}))  \md\bm{s} \right) \\
               &\approx \ln \left[ \sum_j a_{ij} \exp(\beta + \xi(\bm{s}_{ij})) \right].
\end{aligned}
$$
\pkg{inlabru} supports the construction of integration schemes primarily through the \code{fmesher::fm\_int()} function.
This function mainly provides support for defining integration with respect to the finite element mesh for the Mat\'ern SPDE effect.
This integration scheme can be constructed in the following way:
\begin{Schunk}
\begin{Sinput}
R> ips <- fm_int(
+    domain = mesh,
+    samplers = glasgow
+  )
R> 
R> agg <- bru_mapper_logsumexp(
+    n_block = nrow(glasgow),
+    rescale = FALSE
+  )
R> 
R> # For explicit construction of the aggregation weight matrix:
R> W <- fm_block(
+    block = ips$.block,
+    weights = ips$weight,
+    rescale = FALSE
+  )
\end{Sinput}
\end{Schunk}
By default, the integration scheme uses weights at each mesh node, but higher resolution integration schemes can also be constructed.
The \code{fm\_block()} function can be used to constructs the integration matrix by using the \code{block} and \code{weights} arguments.  \code{rescale = FALSE} indicates that the weights in each polygon should not be rescaled to sum to one. The \code{bru\_mapper\_logsumexp()} function allows a more numerically stable implementation of the aggregation calculations,
using internal weight shifts to avoid potential numerical over/underflow.

\section{Approximation Accuracy}
\label{sec:supplement:accuracy}

Section~\ref{sec-approximation-accuracy} of the paper discussed two options
for assessing the accuracy of the linearised model estimates.  We now present
the details required for the Kullback-Leibler divergence diagnostics, that compares
the posterior conditional densities $\wt{p}(\bm{u} | \bm{y}, \bm{\theta} )$ and $\ol{p}(\bm{u} |\bm{y},\bm{\theta})$, evaluated at the posterior mode for $\bm{\theta}$.
With Bayes' theorem,
$$
\begin{aligned}
    p(\bm{u}|\bm{y},{\bm{\theta}}) &= \frac{p(\bm{u},\bm{y}|{\bm{\theta}})}{p(\bm{y}|{\bm{\theta}})} \\
    &= \frac{p(\bm{y}|\bm{u},{\bm{\theta}}) p(\bm{u}|{\bm{\theta}})}{p(\bm{y}|{\bm{\theta}})},
\end{aligned}
$$
where $p(\bm{u}|\bm{\theta})$ is a Gaussian density and $p(\bm{y}|\bm{\theta})$ is a normalisation factor.

\begin{theorem}
Let $\pN(\ol{\bm{m}},\ol{\bm{Q}}^{-1})$ be the linearised posterior distribution
for $\bm{\theta}=\wh{\bm{\theta}}$.
Let
$$
g_i^*=\left.\frac{\partial}{\partial\eta_i} \ln p (\bm{y} | {\bm{\theta}}, \bm{\eta}) \right|_{\bm{\eta}^*}
$$
and
$$
\bm{H}^*_i = \left.\nabla_{u}\nabla_{u}^\top\wt{\eta}_i(\bm{u})\right|_{\bm{u}_*} .
$$
and form the sum of their products, $\bm{G}=\sum_i g_i^*\bm{H}_i^*$.

A Taylor approximation for the difference in observation log-densities between the linearised and original models leads to an approximate non-linear posterior distribution $\pN(\wt{\bm{m}},\wt{\bm{Q}}^{-1})$, where
\begin{align*}
\wt{\bm{Q}} &= \ol{\bm{Q}} - \bm{G}, \\
\wt{\bm{Q}}\wt{\bm{m}} &= \ol{\bm{Q}}\ol{\bm{m}} - \bm{G}\bm{u}_* .
\end{align*}

The K-L divergences are given by
\begin{align*}
\pKL{\ol{p}}{\wt{p}} &= \frac{1}{2}
\left[
\ln\det(\ol{\bm{Q}})-
\ln\det(\ol{\bm{Q}}-\bm{G})
+\tr\left(\ol{\bm{Q}}-\bm{G})\ol{\bm{Q}}^{-1}\right)
-d
+
(\ol{\bm{m}}-\wt{\bm{m}})^\top(\ol{\bm{Q}}-\bm{G})(\ol{\bm{m}}-\wt{\bm{m}})
\right] ,
\\
\pKL{\wt{p}}{\ol{p}} &= \frac{1}{2}
\left[
\ln\det(\ol{\bm{Q}}-\bm{G})-
\ln\det(\ol{\bm{Q}})
+\tr\left(\ol{\bm{Q}}(\ol{\bm{Q}}-\bm{G})^{-1}\right)
-d
+
(\ol{\bm{m}}-\wt{\bm{m}})^\top \ol{\bm{Q}} (\ol{\bm{m}}-\wt{\bm{m}})
\right] .
\end{align*}
\end{theorem}

\subsubsection{Proof of Theorem 1}

\begin{proof}
Recall that the observation likelihood only depends on $\bm{u}$ through $\bm{\eta}$.
Using a Taylor expansion with respect to $\bm{\eta}$ and $\bm{\eta}^*=\wt{\bm{\eta}}(\bm{u}_*)$,
$$
\begin{aligned}
    \ln p(\bm{y}|\bm{\eta},\bm{\theta}) &=
    \ln p (\bm{y}|{\bm{\theta}},\bm{\eta}^*))  \\
    &\qquad + \sum_i \left.\frac{\partial}{\partial\eta_i} \ln p (\bm{y} | {\bm{\theta}}, \bm{\eta}) \right|_{\bm{\eta}^*}\cdot (\eta_i - \eta^*_i) \\
    &\qquad + \frac{1}{2}\sum_{i,j} \left.\frac{\partial^2}{\partial\eta_i\partial\eta_j} \ln p (\bm{y} | {\bm{\theta}}, \bm{\eta}) \right|_{\bm{\eta}^*}\cdot (\eta_i - \eta^*_i) (\eta_j - \eta^*_j) + \mathcal{O}(\|\bm{\eta}-\bm{\eta}^*\|^3),
\end{aligned}
$$
Similarly, for each component of $\wt{\bm{\eta}}$,
$$
\begin{aligned}
\wt{\eta}_i(\bm{u}) &= \eta^*_i
+\left[\left.\nabla_{u}\wt{\eta}_i(\bm{u})\right|_{\bm{u}_*}\right]^\top (\bm{u} - \bm{u}_*)
\\&\quad
+\frac{1}{2}(\bm{u} - \bm{u}_*)^\top\left[\left.\nabla_{u}\nabla_{u}^\top\wt{\eta}_i(\bm{u})\right|_{\bm{u}_*}\right] (\bm{u} - \bm{u}_*) + \mathcal{O}(\|\bm{u}-\bm{u}_*\|^3)
\\&= \eta_i^* + b_i(\bm{u}) + h_i(\bm{u}) + \mathcal{O}(\|\bm{u}-\bm{u}_*\|^3)
\\&= \ol{\eta}_i(\bm{u}) + h_i(\bm{u}) + \mathcal{O}(\|\bm{u}-\bm{u}_*\|^3)
\end{aligned}
$$
where $\nabla_u\nabla_u^\top$ is the Hessian with respect to $\bm{u}$, $b_i$ are linear in $\bm{u}$, and $h_i$ are quadratic in $\bm{u}$.
Combining the two expansions and taking the difference between the full and linearised log-likelihoods, we get
$$
\begin{aligned}
    \ln \wt{p}(\bm{y}|\bm{u},\bm{\theta}) -
    \ln \ol{p}(\bm{y}|\bm{u},\bm{\theta})
    &=
    \sum_i \left.\frac{\partial}{\partial\eta_i} \ln p (\bm{y} | {\bm{\theta}}, \bm{\eta}) \right|_{\bm{\eta}^*}\cdot h_i(\bm{u}) + \mathcal{O}(\|\bm{u}-\bm{u}_*\|^3)
\end{aligned}
$$
Note that the log-likelihood Hessian difference contribution only involves third order $\bm{u}$ terms
and higher, so the expression above includes all terms up to second order.

Let\footnote{Note: this step requires evaluation of the gradient at $\bm{\eta}_*$ which is the latent mode for the hyperparameter mode $\wh{\bm{\theta}}$, and this gradient is currently not evaluated for other $\bm{\theta}$ values.}
\begin{align*}
g_i^*=\left.\frac{\partial}{\partial\eta_i} \ln p (\bm{y} | {\bm{\theta}}, \bm{\eta}) \right|_{\bm{\eta}^*}
&\qquad\text{and}\qquad
\bm{H}^*_i = \left.\nabla_{u}\nabla_{u}^\top\wt{\eta}_i(\bm{u})\right|_{\bm{u}_*} ,
\end{align*}
and form the sum of their products, $\bm{G}=\sum_i g_i^*\bm{H}_i^*$.
Then
\begin{align}
    \ln \wt{p}(\bm{y}|\bm{u},\bm{\theta}) -
    \ln \ol{p}(\bm{y}|\bm{u},\bm{\theta})
    &=
    \frac{1}{2}
    \sum_i g_i^* (\bm{u}-\bm{u}_*)^\top \bm{H}_i^* (\bm{u}-\bm{u}_*) + \mathcal{O}(\|\bm{u}-\bm{u}_*\|^3)
    \\&=
    \frac{1}{2}
    (\bm{u}-\bm{u}_*)^\top \bm{G} (\bm{u}-\bm{u}_*) + \mathcal{O}(\|\bm{u}-\bm{u}_*\|^3).
    \label{eq:observation-discrepancy}
\end{align}
With $\ol{\bm{m}}_{\bm{\theta}}=\pE_{\ol{p}}(\bm{u}|\bm{y},\bm{\theta})$ and
$\ol{\bm{Q}}^{-1}_{\bm{\theta}}=\pCov_{\ol{p}}(\bm{u},\bm{u}|\bm{y},\bm{\theta})$, we obtain
\begin{align}
\pE_{\ol{p}}\left[ \nabla_{\bm{u}}  \left\{\ln \wt{p}(\bm{y}|\bm{u},\bm{\theta}) -
    \ln \ol{p}(\bm{y}|\bm{u},\bm{\theta})\right\}\right]
 &=
    \bm{G}(\ol{\bm{m}}_{\bm{\theta}}-\bm{u}_*)
    + \mathcal{O}\left(\pE_{\ol{p}}\left[\|\bm{u}-\bm{u}_*\|^2\middle|\bm{y},\bm{\theta}\right]\right)
    ,
    \\
\pE_{\ol{p}}\left[ \nabla_{\bm{u}}\nabla_{\bm{u}}^\top  \left\{\ln \wt{p}(\bm{y}|\bm{u},\bm{\theta}) -
    \ln \ol{p}(\bm{y}|\bm{u},\bm{\theta})\right\}\right]
 &=
    \bm{G}
        + \mathcal{O}\left(\pE_{\ol{p}}\left[\|\bm{u}-\bm{u}_*\|\middle|\bm{y},\bm{\theta}\right]\right)
,
    \\
\pE_{\ol{p}}\left[    \ln \wt{p}(\bm{y}|\bm{u},\bm{\theta}) -
    \ln \ol{p}(\bm{y}|\bm{u},\bm{\theta})\right]
    \nonumber
 &=
    \frac{1}{2}
    \tr(\bm{G}\ol{\bm{Q}}_{\bm{\theta}}^{-1}) + \frac{1}{2} (\ol{\bm{m}}_{\bm{\theta}}-\bm{u}_*)^\top\bm{G}(\ol{\bm{m}}_{\bm{\theta}}-\bm{u}_*) \\
&\phantom{= } + \mathcal{O}\left(\pE_{\ol{p}}\left[\|\bm{u}-\bm{u}_*\|^3\middle|\bm{y},\bm{\theta}\right]\right)
.
\label{eq:neg-y-KL-div1}
\end{align}
For each $\bm{\theta}$ configuration in the INLA output, we can extract both $\ol{\bm{m}}_{\bm{\theta}}$ and the sparse precision matrix $\ol{\bm{Q}}_{\bm{\theta}}$ for the Gaussian approximation. The non-sparsity structure of $\bm{G}$ is contained in the non-sparsity of $\ol{\bm{Q}}_{\bm{\theta}}$, which allows the use of Takahashi recursion (implemented by \code{INLA::inla.qinv(Q)}) to compute the corresponding $\ol{\bm{Q}}^{-1}_{\bm{\theta}}$ values needed to evaluate the trace $\tr(\bm{G}\ol{\bm{Q}}^{-1}_{\bm{\theta}})$.
Thus, to implement a numerical approximation of this error analysis only needs special access to the log-likelihood derivatives $g_i^*$, as $\bm{H}_i^*$ can in principle be evaluated numerically.

For a given $\bm{\theta}$,
$$
\begin{aligned}
\pKL{\ol{p}}{\wt{p}} &=
E_{\ol{p}}\left[\ln\frac{\ol{p}(\bm{u}|\bm{y},\bm{\theta})}{\wt{p}(\bm{u}|\bm{y},\bm{\theta})}\right]
\\&=
E_{\ol{p}}\left[
\ln\frac{\ol{p}(\bm{y}|\bm{u},\bm{\theta})}{\wt{p}(\bm{y}|\bm{u},\bm{\theta})}
\right]
-
\ln\frac{\ol{p}(\bm{y}|\bm{\theta})}{\wt{p}(\bm{y}|\bm{\theta})} .
\end{aligned}
$$
The first term can be approximated via the negative of \eqref{eq:neg-y-KL-div1}. For the second term,
\begin{align*}
\ln
\frac{\ol{p}(\bm{y}|\bm{\theta})}{\wt{p}(\bm{y}|\bm{\theta})} &=
\ln
\frac{\ol{p}(\bm{y}|\bm{\theta}) \ol{p}(\bm{u}_*|\bm{\theta},\bm{y})}{\ol{p}(\bm{u}_*|\bm{\theta},\bm{y})}
-
\ln
\frac{\wt{p}(\bm{y}|\bm{\theta}) \wt{p}(\bm{u}_*|\bm{\theta},\bm{y})}{\wt{p}(\bm{u}_*|\bm{\theta},\bm{y})}
\\&=
\ln
\frac{\ol{p}(\bm{y},\bm{u}_*|\bm{\theta})}{\ol{p}(\bm{u}_*|\bm{\theta},\bm{y})}
-
\ln
\frac{\wt{p}(\bm{y},\bm{u}_*|\bm{\theta})}{\wt{p}(\bm{u}_*|\bm{\theta},\bm{y})}
\\&=
\ln
\frac{\ol{p}(\bm{y}|\bm{\theta},\bm{u}_*) p(\bm{u}_*|\bm{\theta})}{\ol{p}(\bm{u}_*|\bm{\theta},\bm{y})}
-
\ln
\frac{\wt{p}(\bm{y}|\bm{\theta},\bm{u}_*) p(\bm{u}_*|\bm{\theta})}{\wt{p}(\bm{u}_*|\bm{\theta},\bm{y})}
\\&=
\ln
\frac{\ol{p}(\bm{y}|\bm{\theta},\bm{u}_*)}{\ol{p}(\bm{u}_*|\bm{\theta},\bm{y})}
-
\ln
\frac{\wt{p}(\bm{y}|\bm{\theta},\bm{u}_*)}{\wt{p}(\bm{u}_*|\bm{\theta},\bm{y})}
\\&=
\ln
\frac{\ol{p}(\bm{y}|\bm{\theta},\bm{u}_*)}{\wt{p}(\bm{y}|\bm{\theta},\bm{u}_*)}
-
\ln
\frac{\ol{p}(\bm{u}_*|\bm{\theta},\bm{y})}{\wt{p}(\bm{u}_*|\bm{\theta},\bm{y})} .
\\&=
-
\ln
\frac{\ol{p}(\bm{u}_*|\bm{\theta},\bm{y})}{\wt{p}(\bm{u}_*|\bm{\theta},\bm{y})} ,
\end{align*}
where the last step uses that both observation densities are evaluated at the linearisation point, so that the predictors are identical, so the two observation densities are the same.
Now, by approximating both the linearised and non-linearised posterior distributions with Gaussian distributions $\pN(\ol{\bm{m}}_{\bm{\theta}},\ol{\bm{Q}}^{-1}_{\bm{\theta}})$
and $\pN(\wt{\bm{m}}_{\bm{\theta}},\wt{\bm{Q}}^{-1}_{\bm{\theta}})$, respectively,
the observation log-density discrepancy \eqref{eq:observation-discrepancy}
gives a approximate equation, for some constant $C$,
\begin{align*}
\ln \wt{p}_G(\bm{u}|\bm{\theta},\bm{y}) -
\ln \ol{p}_G(\bm{u}|\bm{\theta},\bm{y}) &=
\ln \wt{p}(\bm{y}|\bm{\theta},\bm{u}) + \ln p(\bm{u}|\bm{\theta})-
\ln \ol{p}(\bm{y}|\bm{\theta},\bm{u}) - \ln p(\bm{u}|\bm{\theta}) + C
\\&\approx
\frac{1}{2}(\bm{u}-\bm{u}_*)^\top\bm{G}(\bm{u}-\bm{u}_*) + C .
\end{align*}
In order for the quadratic expressions in $\bm{u}$ to match between the left and right hand sides,
we need $-\wt{\bm{Q}}_{\bm{\theta}}+\ol{\bm{Q}}_{\bm{\theta}}=\bm{G}$ and $\ol{\bm{Q}}_{\bm{\theta}}(\wt{\bm{m}}_{\bm{\theta}}-\ol{\bm{m}}_{\bm{\theta}})=\bm{G}(\wt{\bm{m}}_{\bm{\theta}}-\bm{u}_*)$, or equivalently, $\wt{\bm{Q}}_{\bm{\theta}}=\ol{\bm{Q}}_{\bm{\theta}}-\bm{G}$ and $\wt{\bm{Q}}\wt{\bm{m}}_{\bm{\theta}}=\ol{\bm{Q}}_{\bm{\theta}}\ol{\bm{m}}_{\bm{\theta}}-\bm{G}\bm{u}_*$.
The K-L divergence for a given $\bm{\theta}$ becomes
\begin{align*}
\nonumber
\pKL{\ol{p}}{\wt{p}} &\approx
\frac{1}{2}
\left[
\ln\det(\ol{\bm{Q}}_{\bm{\theta}})-
\ln\det(\ol{\bm{Q}}_{\bm{\theta}}-\bm{G})
-\tr\left(\bm{G}\ol{\bm{Q}}^{-1}_{\bm{\theta}}\right)
+
(\ol{\bm{m}}_{\bm{\theta}}-\bm{u}_*)^\top\bm{G}(\ol{\bm{Q}}_{\bm{\theta}}-\bm{G})^{-1}\bm{G}(\bm{m}_{\bm{\theta}}-\bm{u}_*)
\right] .
\end{align*}
where $\bm{Q}$ and $\bm{G}$ are known matrices, $\bm{u}_*$ is the linearisation location and $\bm{m}=\pE_{\ol{p}}(\bm{u}|\bm{y},\bm{\theta})$.

Using the relationships between $\wt{\bm{Q}}_{\bm{\theta}}$, $\wt{\bm{m}}_{\bm{\theta}}$, $\ol{\bm{Q}}_{\bm{\theta}}$, $\ol{\bm{m}}_{\bm{\theta}}$, and $\bm{u}_*$, we can rewrite this as
\begin{align*}
\nonumber
\pKL{\ol{p}}{\wt{p}} &\approx
\frac{1}{2}
\left[
\ln\det(\ol{\bm{Q}}_{\bm{\theta}})-
\ln\det(\ol{\bm{Q}}_{\bm{\theta}}-\bm{G})
-\tr\left(\bm{G}\ol{\bm{Q}}^{-1}_{\bm{\theta}}\right)
+
(\wt{\bm{m}}_{\bm{\theta}}-\ol{\bm{m}}_{\bm{\theta}})^\top(\ol{\bm{Q}}_{\bm{\theta}}-\bm{G})(\wt{\bm{m}}_{\bm{\theta}}-\ol{\bm{m}}_{\bm{\theta}})
\right] .
\end{align*}

\end{proof}

\section{Iterative inlabru verbose output for Example 5.3}
\label{sec:supplement:verbose}

The amount of information printed by the \code{bru()} function can be set by the \code{bru\_verbose} option, which has 4 levels of increasing verbosity.  This information can also be accessed from the output object at a later time, as shown below, including higher verbosity levels.

\begin{Schunk}
\begin{Sinput}
R> fit_3 <- bru(
+       components = cmp,
+       z_lik,
+       count_lik,
+       options = list(
+         bru_verbose = 1,
+       )
+     )
R> print(bru_log(fit_3, verbosity = 2), timestamp = FALSE, verbosity = TRUE)
\end{Sinput}
\end{Schunk}
\footnotesize
\begin{Schunk}
\begin{Soutput}
iinla: Iteration 1 [max:10] (level 1)
iinla: Iteration 2 [max:10] (level 1)
iinla: Step rescaling: 99.5% (norm0 = 94.74, norm1 = 5.154, norm01 = 95.16) (level 2)
iinla: Max deviation from previous: 1500% of SD, and line search is active
       [stop if: <10% and line search inactive] (level 1)
iinla: Iteration 3 [max:10] (level 1)
iinla: Step rescaling: 97.6% (norm0 = 4.934, norm1 = 0.1598, norm01 = 4.941) (level 2)
iinla: Max deviation from previous: 108% of SD, and line search is active
       [stop if: <10% and line search inactive] (level 1)
iinla: Iteration 4 [max:10] (level 1)
iinla: Step rescaling: 101% (norm0 = 2.613, norm1 = 0.03321, norm01 = 2.614) (level 2)
iinla: Max deviation from previous: 81.7% of SD, and line search is active
       [stop if: <10% and line search inactive] (level 1)
iinla: Iteration 5 [max:10] (level 1)
iinla: Step rescaling: 99.8% (norm0 = 0.2526, norm1 = 0.0007458, norm01 = 0.2526) (level 2)
iinla: Max deviation from previous: 5.74% of SD, and line search is active
       [stop if: <10% and line search inactive] (level 1)
iinla: Iteration 6 [max:10] (level 1)
iinla: Max deviation from previous: 4.28% of SD, and line search is inactive
       [stop if: <10% and line search inactive] (level 1)
iinla: Convergence criterion met.
       Running final INLA integration step with known theta mode. (level 1)
iinla: Iteration 7 [max:10] (level 1)
\end{Soutput}
\end{Schunk}
\normalsize

\end{document}